\documentclass{SciPost}

\newcommand{\vphi}{\varphi}
\newcommand{\ra}{\rightarrow}
\newcommand{\crat}{\zeta}

\newcommand{\mc}[1]{\mathcal{#1}}

\newcommand{\m}{\mathbf{m}}

\newcommand{\ov}[1]{\overline{#1}}

\newcommand{\mief}{\mathrm{MIE_F}}

\hypersetup{
    colorlinks,
    linkcolor={red!50!black},
    citecolor={blue!50!black},
    urlcolor={blue!80!black},
    pdftitle={MIEdistributiondraft}
}
\usepackage{xcolor}
\definecolor{myteal}{HTML}{0F6E6C} 

\usepackage{physics}
\usepackage{bm}
\usepackage{ulem}
\usepackage{graphicx}
\usepackage[bitstream-charter]{mathdesign}
\usepackage{mathtools}
\usepackage{comment}
\usepackage{subcaption} 
\allowdisplaybreaks
\DeclareSymbolFont{usualmathcal}{OMS}{cmsy}{m}{n}
\DeclareSymbolFontAlphabet{\mathcal}{usualmathcal}

\fancypagestyle{SPstyle}{
\fancyhf{}
\lhead{\colorbox{scipostblue}{\bf \color{white} ~SciPost Physics}}
\rhead{{\bf \color{scipostdeepblue} ~Submission }}

\fancyfoot[C]{\textbf{\thepage}}
}
\DeclareUnicodeCharacter{2009}{\,}
\usepackage{amsmath}
\usepackage{cancel}
\begin{document}

\pagestyle{SPstyle}

\begin{center}{\Large \textbf{\color{scipostdeepblue}{
Universal Statistics of Measurement-Induced Entanglement in Tomonaga-Luttinger liquids
}}}\end{center}

\begin{center}\textbf{
Kabir Khanna\textsuperscript{$\star$},
Romain Vasseur\textsuperscript{$\dagger$}
}\end{center}

\newcommand{\umass}{Department of Physics, University of Massachusetts, Amherst, MA, USA}
\newcommand{\unige}{ Department of Theoretical Physics, University of Geneva, 24 quai Ernest-Ansermet, 1211 Gen\`eve, Switzerland}
\begin{center}
\unige\\
$\star$ \href{mailto:email1}{\small kabir.khanna@unige.ch}\,,\quad
$\dagger$ \href{mailto:email2}{\small romain.vasseur@unige.ch}
\section*{\color{scipostdeepblue}{Abstract}}
\end{center}
\textbf{\boldmath{%
We study the statistics of measurement-induced entanglement (MIE) after partial measurement on a class of one-dimensional quantum critical states described by Tomonaga-Luttinger liquids at low energies. Using a replica trick to average over measurement outcomes in the charge basis and tools from conformal field theory (CFT), we derive closed-form expressions for the cumulants of MIE. We show that exact Born-averaging over microscopic measurement outcomes becomes equivalent at low energy to averaging over conformal boundary conditions weighted by their corresponding partition functions. Our results yield distinctive critical behavior across all cumulants in the regime where the unmeasured parts of the system are maximally separated. We also obtain the full distribution of MIE, finding that it is generically bimodal and exhibits fat-tails. We corroborate our analytical predictions by numerical calculations and find good agreement between them. 
}}

\vspace{\baselineskip}



\vspace{10pt}
\noindent\rule{\textwidth}{1pt}
\tableofcontents
\noindent\rule{\textwidth}{1pt}
\vspace{10pt}


\section{Introduction}
\label{sec:intro}
The past decade has seen an increasing interest in understanding the effects of measurements on many-body systems, at equilibrium or otherwise. Such an interest is not unprompted: measurements have been a key to some fundamental quantum information and computational protocols such as teleportation \cite{bennett1993teleporting}, error-correction \cite{PhysRevA.52.R2493, gottesman1997stabilizer}, and measurement-based quantum computation (MBQC) \cite{raussendorf2003measurement}. In many-body physics, the broader pursuit of realizing physical systems capable of quantum computation (usually ground states and low-lying excitations of systems with topological order)  has similarly highlighted the utility of measurements while also revealing its non-trivial effects on many-body states. In particular, although measurements are typically viewed as disentangling operations, selectively measuring only part of a many-body system can have highly non-trivial effects on the unmeasured degrees of freedom, such as reshaping their entanglement structure. Several works \cite{PhysRevLett.127.220503, PhysRevX.14.021040, verresen2021efficiently, PRXQuantum.3.040337, iqbal_topological_2024, foss2023experimental, iqbal_non-abelian_2024, bluvstein_logical_2024} exploit precisely this feature, using measurements as a ``shortcut'' to efficiently prepare long-range entangled topological states that could be used for quantum computation, a procedure that would otherwise naively require macroscopically deep unitary circuits. From a many-body perspective, such a computational way of thinking of states using their preparation methods provides a novel route to analyze and classify phases of matter \cite{PhysRevLett.127.220503,PRXQuantum.4.020339}. Meanwhile, studies in non-equilibrium systems uncovered a fundamental dynamical phase transition driven by measurements, namely, measurement-induced phase transitions (MIPTs) \cite{PhysRevX.9.031009, PhysRevB.98.205136, PhysRevB.100.134306, PhysRevB.101.104301,PhysRevB.101.104302,PhysRevX.10.041020,PhysRevLett.125.030505,PhysRevLett.125.070606,PhysRevX.12.041002,PhysRevLett.129.200602, PhysRevB.101.060301,noel_measurement-induced_2022, koh_measurement-induced_2023, hoke_measurement-induced_2023, annurev:/content/journals/10.1146/annurev-conmatphys-031720-030658, Potter2022}, highlighting a dynamical competition between entangling unitary dynamics and disentangling measurements. More recently, measurements have also been used to give a stricter notion of thermalization, dubbed ``deep thermalization'' \cite{choi_preparing_2023, PRXQuantum.4.010311, PhysRevLett.128.060601, PhysRevX.14.041051, PhysRevLett.133.260401, PRXQuantum.6.020343,Ippoliti2022solvablemodelofdeep, PRXQuantum.4.030322, v8kp-39ry, Claeys2022emergentquantum, PhysRevB.108.104317, Varikuti2024unravelingemergence, 10.21468/SciPostPhys.18.3.107, zhang_holographic_2025, lami2025quantumstatedesignemergent}, revealing a novel universal characterization of equilibration in many-body systems. These developments have further been accompanied by their respective experimental investigations on various quantum computational platforms \cite{noel_measurement-induced_2022,choi_preparing_2023, koh_measurement-induced_2023, hoke_measurement-induced_2023, iqbal_topological_2024, foss2023experimental, iqbal_non-abelian_2024, bluvstein_logical_2024}. 

As emphasized in the works discussed above, extracting genuine measurement-induced physics from a partially measured many-body state requires explicitly keeping track of the measurement outcomes and their associated Born probabilities. For example, as one tunes the measurement rate in a generic (non-integrable) unitary evolution interspersed with measurements, MIPTs manifest in the entanglement structure of a typical many-body state \textit{conditioned} on the measurement outcomes. If, by contrast, one does not register these outcomes, the resultant state is effectively fully de-phased over the measurement records, thereby washing out any interesting measurement-related physics and hiding the physics of MIPTs. Likewise, deep thermalization is formulated in terms of the \textit{projected ensemble} \cite{choi_preparing_2023, Claeys2022emergentquantum}: an ensemble of pure states on the unmeasured region, conditioned on and weighted by the outcomes of projective measurements performed in the complementary region. Consequently, observables that probe such measurement-induced phenomena typically involve averages over measurement outcomes weighted by their Born probabilities. Moreover, they must be non-linear functionals of the density matrix; otherwise, they coincide with the same observable evaluated in the state that is fully dephased over the measurement outcomes. From a theoretical standpoint, computing such observables is challenging, as it requires detailed knowledge of the many-body Born probabilities and an average over an exponentially large space of measurement records.
 
Some analytic progress can nevertheless be made, for example in MIPTs, by averaging over all possible evolutions to extract universal physics. But even in those cases, one often uses the replica trick and finally takes a replica limit \cite{PhysRevB.101.104301, PhysRevB.101.104302} to retrieve the true measurement averaged physics; a task that has proven challenging and is possible only in special limits \cite{PhysRevB.101.104301, PhysRevB.101.104302} and non-trivially in certain cases \cite{PhysRevX.13.041045}, often requiring non-local setups \cite{de2023universality,giachetti2023elusive, bulchandani_random-matrix_2024}. As a result, most existing theoretical works use physics extracted without taking the replica limit (for example, multi-replica physics) \cite{BAO2021168618, PhysRevB.108.104310, PhysRevA.109.042414}, or forcing measurements (post-selection) to specific outcomes as proxies for a true measurement average \cite{PhysRevB.92.075108, Rajabpour_2016, najafi_entanglement_2016, PhysRevB.111.155143}. Experimentally however, it is important to note that such measurement-averaged observables are hard to measure: they require one to prepare the same post-measurement state multiple times, an event that is exponentially unlikely in the system size. Despite this, these observables indeed hold operational meaning making them worthy of analytic investigation. For example, the entanglement transition of MIPTs that is probed by the typical entanglement structure coincides with a transition in the error correction capabilities of chaotic unitary dynamics \cite{PhysRevLett.125.030505}. 

A particularly important post-measurement observable underlying measurement-related phenomena is the measurement-induced entanglement (MIE) \cite{Lin2023probingsign, PhysRevLett.92.027901, PhysRevA.71.042306}, which quantifies the average entanglement between two regions after the remainder of the system has been locally measured. Concretely, for a chosen (unmeasured) region $A$, MIE is defined as the entanglement entropy of $A$ in the post-measurement state, averaged over all measurement outcomes with their respective Born weights. A closely related quantity first appeared in the context of localizable entanglement \cite{PhysRevLett.92.027901, PhysRevA.71.042306}, where it was used to bound two-point correlations. Since then, MIE has proven operationally useful in a variety of settings, including measurement-based quantum computation (MBQC), diagnosing sign problems in many-body states \cite{Lin2023probingsign, hastings_how_2015}, probing the complexity of tensor network contractions \cite{PhysRevResearch.3.033002, chertkov_holographic_2022} and sampling tasks in random quantum circuits \cite{PhysRevX.12.021021, PhysRevX.15.021059,PhysRevLett.132.030401,watts2024quantumadvantagemeasurementinducedentanglement} where it was proposed to be a probe for quantum advantage \cite{PRXQuantum.6.010356}, detecting teleportation transitions \cite{PhysRevLett.132.030401}, and bounding strange correlators \cite{PhysRevLett.112.247202}. 

Besides its quantum information-based applications, MIE, much like entanglement entropy, also serves as a probe of quantum phases \cite{Lin2023probingsign, PhysRevB.109.195128}. Restricting to 1+1D, for generic gapped phases MIE decays exponentially with the separation between the unmeasured regions \cite{hastings_how_2015}, whereas in symmetry-protected topological phases it remains nonzero and constant when measurements are performed in a symmetry-preserving basis \cite{PhysRevLett.108.240505}. Remarkably, for ground states of 1+1D quantum critical systems, numerical evidence \cite{Lin2023probingsign, PhysRevB.109.195128} has suggested that MIE is \textit{conformally invariant} and apparently universal, exhibiting novel critical exponents distinct from those obtained in approaches based on post-selected outcomes \cite{Rajabpour_2016, najafi_entanglement_2016}. However, a general analytic understanding of these features, derived from a genuine measurement average over Born probabilities, has remained elusive precisely because of the theoretical difficulties outlined above. In our recent work \cite{khanna2025measurementinducedentanglementconformalfield}, we addressed this challenge by deriving an exact analytic expression for MIE in a broad class of 1+1D quantum critical systems governed at low energies by the free-boson CFT, commonly known as Tomonaga–Luttinger liquids (TLLs) \cite{tomonaga_remarks_1950, luttinger_exactly_1963,F_D_M_Haldane_1981,PhysRevLett.47.1840,  giamarchi_quantum_2003}. Specifically, we considered projective measurements of the charge density in the geometry shown in Fig.~\ref{Fig1} and computed the entanglement entropy of a region $A$, averaged over all measurement outcomes with their Born probabilities. This provided a rare example in which the randomness of measurement outcomes could be treated by implementing the replica trick exactly. Our results further established the universality of MIE in the underlying phase and showed that post-selecting to special outcomes forms a poor proxy for genuine Born-averaged effects. Finally, we note that related questions have been explored in a number of works on weak measurements on critical ground states \cite{garratt2023measurements, PhysRevX.13.041042, yang2023entanglement, patil2024highly, PRXQuantum.4.030317, sun2023new}. In those studies, weak measurements are typically applied to the entire system, in contrast to our protocol where only a subregion is projectively measured, but the focus is likewise on post-measurement observables. Methodologically, however, the approaches are quite distinct: the weakness of the measurements in those works permits a controlled perturbative expansion, whereas the projective limit considered here necessitates a fully non-perturbative treatment. Even in those settings, many of the theoretical difficulties discussed above persist whenever the measurement is ``relevant'' in the RG sense, and as mentioned before, these works often have to resort to multi-replica physics or forcing specific outcomes (exceptions include~\cite{patil2024highly}). 

In this work, we further advance our analytic toolbox to study the statistics of MIE in TLLs. The motivation for this exercise is twofold. On the one hand, it allows us to further characterize the universality of MIE in TLLs. On the other, it has a more practical role: it provides a rigorous understanding of whether performing a single-shot measurement is able (or unable) to generate entanglement upon measurement, and more importantly to what extent, aspects that are invisible at the level of a simple average and instead require a full distributional description. Ultimately, MIE is a statistical quantity, which is characterized by a probability distribution, rather than just its mean. In fact, in the case of TLLs, we will show that this distribution is very far from Gaussian, and its mean is not representative of typical outcomes.

To this end, using the replica trick on what we derive to be the generalized replica partition function, we obtain closed-form expressions for all higher cumulants of MIE. These expressions are universal and establish a crucial observation made in our previous work: Born averaging over microscopic outcomes is equivalent to averaging over conformal boundary conditions of the underlying CFT with a well-defined measure. Consequently, MIE and its cumulants equal the average (over conformal boundary conditions) of the “forced” MIE and its cumulants, where the latter are conditioned on a fixed outcome. As a corollary, we use this recipe to also extract entanglement induced by quenched (impurity-like) disorder, a quantity we call disorder induced entanglement (DIE). For MIE, we further extract the critical exponents in the limit of maximal separation between unmeasured regions, finding an identical universal scaling across all cumulants for sufficiently large Rényi index $n$, with qualitative changes at smaller $n$. To further probe this scaling, we analyze the full distribution of MIE and show it is heavy-tailed while being universally bimodal with peaks at both extremes, with the higher-value peak vanishing in the maximal-separation (or small cross-ratio) limit. The scaling of the cumulants of MIE in this limit is governed, as we show, by this higher-value tail of the distribution, corresponding to rare measurement outcomes that generate large amounts of entanglement. Notably, this vanishing peak sits at $\log 2$, implying a vanishingly small yet finite probability to generate an EPR pair, a feature that indicates that the critical state effectively acts as a quantum ``wire'' \cite{PhysRevLett.108.240505, PhysRevLett.119.010504} for certain outcomes. Collectively, these results furnish one of the first complete descriptions, at a deeper level of distributions, of how real measurements reshape the universal entanglement structure in TLLs.

\section{Setup}
We start by defining the setup and the quantities of interest. While our primary interest is in the long-wavelength continuum description of TLLs, it is useful to motivate the setup concretely via a lattice realization. To this end, we consider the spin-1/2 XXZ chain on a ring with Hamiltonian
\begin{equation}
\label{xxzhamiltonian}
H = \sum_i \sigma^x_i\sigma^x_{i+1} + \sigma^y_i\sigma^y_{i+1} + \Delta\sigma^z_i\sigma^z_{i+1},
\end{equation}
which provides a prototypical lattice model whose low-energy physics is described by TLLs in the regime $-1<\Delta\leq 1$. The above model has a $U(1)$ symmetry that is implemented by the unitary $U(\theta) = \prod_{i} e^{i\theta\sigma^z_i},\theta\in (0,2\pi]$, corresponding to global rotations about the $z$-axis. 

\begin{figure}[t!]
    \centering
    \includegraphics[width=\linewidth]{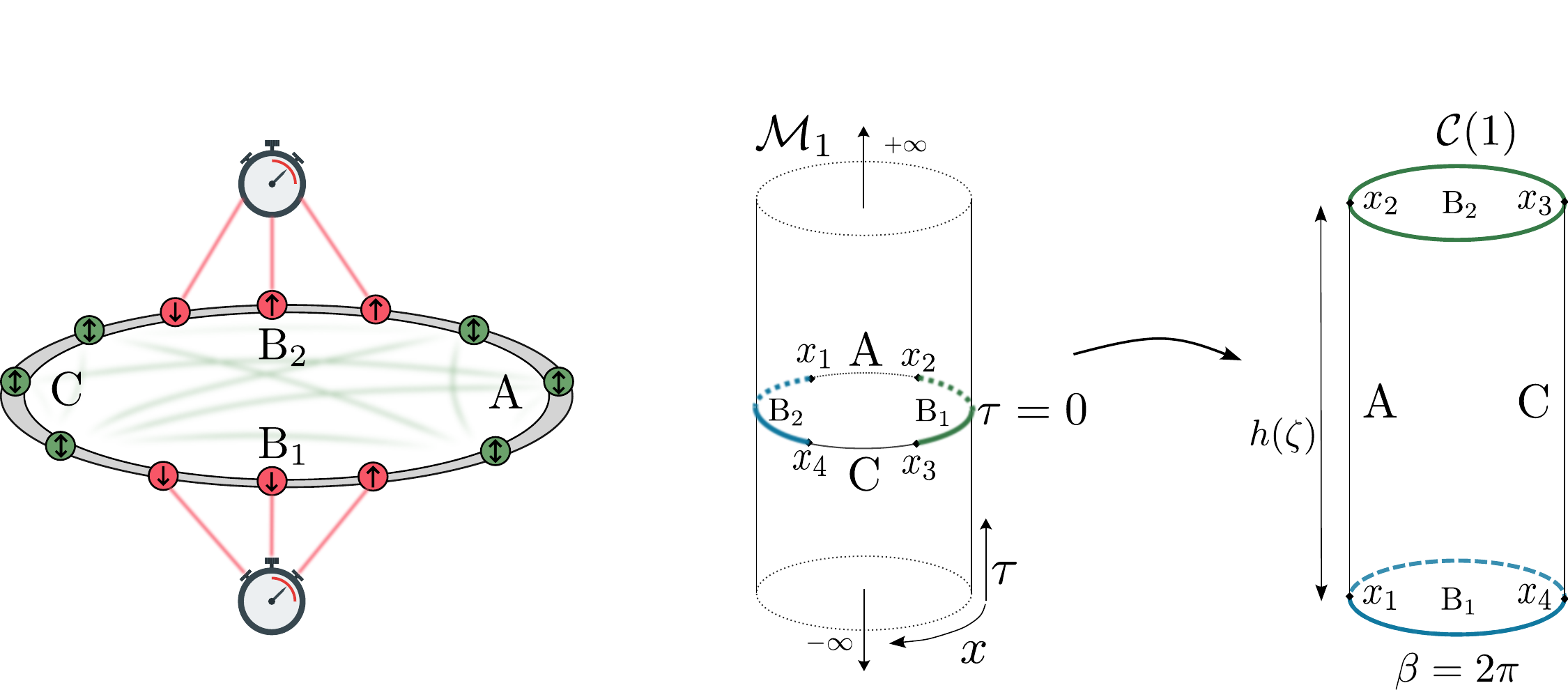}
    \caption{\textbf{Setup. }\textbf{Left}: Schematic illustration of a TLL modeled as a spin-1/2 chain. Red (green) sites denote the measured (unmeasured) sites. Green shaded curves illustrate the entanglement between the unmeasured spins. \textbf{Right}: Manifolds $\mc{M}_1$ and $\mc{C}(1)$, and the conformal mapping between them. The intervals $\mathrm{A} = [x_1,x_2]$ and $\mathrm{C} = [x_3,x_4]$ on the cylinder $\mc M_1$ denote unmeasured regions, while $\mathrm{B} = [x_1,x_4]\cup[x_2,x_3]= \mathrm{B}_1 \cup \mathrm{B}_2$ denotes the measured region. $\mc C(1)$ has circumference $\beta=2\pi$ and length ${h(\zeta)}$ which is a function of the conformal cross ratio $\zeta=w_{12}w_{34}w^{-1}_{13}w_{24}^{-1}$, where $w_{ij} = \frac{L}{\pi} \sin(\pi x_{ij}/L)$.}
    \label{setup} \label{Fig1}
\end{figure}

We consider the (normalized) ground state $\rho_0 = \ket{\psi_0}\bra{\psi_0}$ of the above model and perform single-site projective measurements on a disjoint region $B = B_1\cup B_2$ (see Fig.~\ref{Fig1}). The reason for such a geometry will be made clear shortly. These measurements are carried out in the eigen-basis of the symmetry generator, i.e., the Pauli $\sigma^z$ basis, and the outcomes are labeled via a bit string $\mathbf{m}\in\{0,1\}^{|B|}$, where $|B|$ denotes the length of the measured region $B$. Mathematically, the \textit{un-normalized} post-measurement state conditioned on a \textit{fixed} outcome $\m$ is
\begin{equation}
\label{unnormalizedpostmeasurement}
\rho_\m = \ket{\psi_\m}\bra{\psi_\m}= M_\m\rho_0M_\m,
\end{equation}
where 
\begin{equation}
\label{measurementop}
    M_\m = \prod_{i\in B}\left[\frac{1+(-1)^{\m_i}\sigma^z_i}{2}\right],
\end{equation}
is the relevant measurement operator. The norm of $\rho_\m$ is $\tr(\rho_\m) = p_\m$, which is the probability of obtaining the outcome $\m$. We are now interested in the entanglement entropy of unmeasured sub-regions of the state $\rho_\m$. In particular, consider Born-averaged post-measurement entanglement entropy, also known as MIE, to be 
\begin{equation}
    \mathrm{MIE}(A ) = \ov{S^{(n)}_{\m, A}}=\sum_\m p_\m S^{(n)}_{\m,A},
\end{equation}
where $S^{n}_{\m, A} = (1-n)^{-1} \log(\tr\rho^n_{\m, A}/(\tr\rho_\m)^n)$ with $\tr\rho_{\m, A} = \tr_C(\rho_\m)$. Historically, this quantity has been of operational interest in geometries where the unmeasured regions are separated by a fully measured segment \cite{PhysRevA.71.042306, PhysRevLett.92.027901, PhysRevLett.108.240505, hastings_how_2015, Lin2023probingsign}. This motivates the setup in Fig.~\ref{Fig1}, where we fully measure the region $B$, leaving two disjoint unmeasured regions $A$ and $C$. While our focus throughout this work is primarily on MIE and its cumulants, we will occasionally refer to two related quantities. The first is the forced MIE, denoted $\mathrm{MIE}^{(n)}_{\mathrm F}$, defined as the Rényi entanglement entropy $S^{(n)}_{\m_0,A}$ obtained by fixing, or post-selecting, a particular measurement outcome $\m_0$. The second is the typical MIE, $\mathrm{MIE}^{(n)}_{\mathrm{typical}}=\exp[\sum_{\m}p_{\m}\log S^{(n)}_{\m,A}]$, which suppresses the effect of rare outcomes with large entanglement that can otherwise dominate the Born-averaged MIE and its higher cumulants.

An alternate way to understand MIE is via quantum \textit{conditional mutual information} (CMI), a quantity that is central to mixed-state transitions \cite{PhysRevLett.134.070403}. For a tripartite system ABC, it is defined as $I(A:C|B) = S(AB) + S(BC) -S(B) - S(ABC)$. In other words, it is the mutual information between regions $A$ and $C$ once the state in $B$ is known. The connection with MIE is as follows: If we subject our pure state $\ket{\psi_0}$ (which is supported on ABC) to a measurement channel $\mathcal N[\cdot]$ in region $B$, then the CMI of the resultant state is \cite{PhysRevA.110.032426}
\begin{equation}
    I(A:C|\mathcal{N}[B]) = \sum_{\m} p_\m I(A:C|\m) = 2\mathrm{MIE}(A),
\end{equation}
where the last equation follows from the definition of $I(A:C|\m)$ and the fact that we have a pure state $\ket{\psi_\m}$ post-measurement. 
In a previous work \cite{khanna2025measurementinducedentanglementconformalfield}, we used field-theoretic methods to isolate the universal low-energy contributions to MIE, demonstrating that they are dominant and that non-universal corrections are negligible. In the present work, we go beyond the average and study the full statistics of MIE with $S^{(n)}_{\mathbf{m},A}$ being a random variable over measurement outcomes. In particular, we evaluate its higher moments
\begin{equation}
    \overline{\bigl(S^{(n)}_{\mathbf{m},A}\bigr)^l}
    = \sum_\mathbf{m} p_\mathbf{m}\,\bigl(S^{(n)}_{\mathbf{m},A}\bigr)^l,
\end{equation}
which characterize the resulting probability distribution of $S^{(n)}_{\mathbf{m},A}$ that we also evaluate in this work.\\
In the continuum, at low energies, the gapless XXZ spin-chain is described by a TLL \cite{tomonaga_remarks_1950, luttinger_exactly_1963}. More generally, TLLs arise ubiquitously in one dimension, providing the universal low-energy description of a broad class of metals and interacting bosonic and fermionic systems \cite{giamarchi_quantum_2003}. Their long-wavelength physics is governed by the compact free-boson \cite{F_D_M_Haldane_1981, PhysRevLett.47.1840} CFT with (Euclidean) action \cite{DiFrancesco1997}

\begin{equation}
\label{euclideanaction}
    S=\frac{g}{4\pi}\int dx \int d\tau \left[(\partial_x\vphi)^2 + (\partial_\tau \vphi)^2\right],
\end{equation}
where $g$ is the Luttinger parameter that characterizes correlations in the critical state, and where we have set the Fermi velocity to unity.  The field $\vphi$ is interpreted as the counting field for the $U(1)$ charge in the system and is related to the local charge (density) operator $\hat{n}(x)$ via the bosonization expression \cite{giamarchi_quantum_2003} $\hat n(x)\simeq n_0 +\frac{1}{2\pi}\nabla\hat \vphi(x)+ A\cos(\hat{\vphi}(x)+2\pi n_0x)+\dots$, where $\dots$ represent higher-order oscillations of the counting field that can be ignored at large distances, $A$ is a non-universal factor, and $n_0$ is a reference background (filling fraction) relative to which we track density fluctuations. Crucially for this work, bosonization leads to field $\hat\vphi(x)$ being compact, i.e., $\vphi\sim\vphi+2\pi w$ with radius $r=1$ and an integer winding number $w\in\mathbb{Z}$. 

\section{Main Assumptions and Strategy}
Here we explicitly outline the main assumptions that we make in this work. First, we assume that MIE is conformally invariant, so that it may be computed after mapping the geometry in Fig.~\ref{Fig1} to a more convenient one. Such an assumption is justified from previous numerical studies \cite{Lin2023probingsign, PhysRevB.109.195128} that evaluate MIE on the lattice in the setup shown in Fig.~\ref{Fig1}, finding it to be a function of the cross-ratio $\zeta$. However, we stress that conformal invariance will be utilized only when performing a conformal map and nowhere else in the rest of the calculation. \\Second, 
in the field-theoretic description, we model the projective measurements of the local charge $\sigma^z_j$ in the XXZ chain (or, in the continuum, of the coarse-grained density $\hat n(x)$) as imposing inhomogeneous \textit{random} Dirichlet boundary conditions on the bosonic field $\varphi(x)$ in the measured regions. In other words, at the level of the field theory, we will pin the field $\hat \varphi$ and average over random configurations of this field in the measured region using the replica trick. This replacement is \textit{not} an exact identification of individual lattice outcomes with smooth field configurations. Rather, it is a low-energy field-theoretic prescription for performing the Born-average. For special outcomes, such a correspondence is  well established: for example, the Néel state $\ket{\uparrow\downarrow\uparrow\downarrow\cdots}$, is known to flow to a uniform Dirichlet boundary condition with $\varphi=0$. However, an understanding of the low-energy fate of an arbitrary lattice configuration remains open. This prescription is furthermore natural at the level of symmetries, and in our previous work we found excellent numerical agreement between the MIE computed in the XXZ chain and the continuum prediction obtained by modeling measurement outcomes as random field configurations of the field $\varphi$, providing strong evidence that this replacement is indeed justified at the field-theory level.\\
Finally, we emphasize that our main result Eq.~\eqref{cumulantfinalexpression} will have a natural interpretation in terms of averaging over \textit{uniform} Dirichlet boundary conditions that are indeed also the conformally invariant ones. This however, is only an interpretation of the final result, and at no point in the calculation do we invoke any boundary-CFT formalism to arrive at it. Instead, our calculation starts from random boundary configurations of the field $\varphi(x)$ and shows that the final answer can be re-organized in this simpler form. In summary, our approach should be viewed as a field-theory calculation and not an exact microscopic treatment of the measurement outcomes, which is likely intractable in any case.

\section{Replica calculation}
\label{sectionreplicacalculation}
In this section we present the calculation for higher cumulants of MIE upon measuring the charge operator ($\sigma^z)$ on the XXZ chain in its low energy limit. For simplicity, we begin by considering the mean MIE (or simply, MIE). The first step to evaluate MIE is to use the \textit{replica trick} in order to compute an average over the logarithm in $S^{n}_{\m,A}$:
\begin{align}
     \ov{S^{n}_{\m,A}} := \frac{1}{1-n}\ov{\log\frac{\tr\rho^n_{\m,A}}{(\tr\rho_{\m})^n} }&=\frac{1}{1-n}\lim_{k\ra0} 
      \left[\partial_k[\ov{(\tr\rho^n_{\m,A})^k}] - \partial_k[\ov{(\tr\rho^n_{\m})^k}]\right]  \notag
     \\
     \label{avgmiereplica}
     &=\frac{1}{1-n}\lim_{k\ra0} 
      \left[\partial_k\log[\ov{(\tr\rho^n_{\m,A})^k}] - \partial_k\log[\ov{(\tr\rho^n_{\m})^k}]\right],
\end{align}
where $\rho_\m$ is the un-normalized post-measurement density matrix defined in Eq.~\eqref{unnormalizedpostmeasurement}, $\ov{[{\dots}]}$ denotes an average over the measurement outcomes $\m$ with Born probability $p_\m = \tr \rho_\m$, and where we have used the fact that $\sum_\m p_\m = 1$ in the last equation to introduce logarithms (for a reason that will be clear in a moment). While the replica trick was originally introduced in the context of spin glasses \cite{doi:10.1142/0271} and is useful in dealing with quenched disorder, its specific usage to deal with measurement-related disorder is inspired from the field of measurement-induced criticality~\cite{PhysRevB.100.134203, PhysRevB.101.104301, PhysRevB.101.104302,PhysRevB.99.174205}.

To make further progress, we express $\tr \rho_\m$ in its path-integral representation. This follows from the standard imaginary-time representation of the ground-state density matrix,
$\rho_0 = \ket{\psi_0}\bra{\psi_0} \propto \lim_{\tau\to\infty} e^{-\tau H}\ket{\psi_{\rm init}}\bra{\psi_{\rm init}}e^{-\tau H}$, where $\ket{\psi_{\rm init}}$ is some initial quantum state on our Hilbert space, $H$ is the Hamiltonian, $\tau$ is the Euclidean time, and the limit $\tau \to \infty$ suppresses all excited states relative to the ground state. Introducing continuum quantum fields $\hat{\vphi}(x,\tau)$, with $x$ denoting space and $\tau$ imaginary time, we can write the partition function as
$ Z =\Tr \rho_0 = \int \mc D[\vphi] \exp[-S(\vphi)]$, where $S$ is the Euclidean action from Eq.~\eqref{euclideanaction} and the trace enforces periodic boundary conditions in imaginary time at $\tau=0$ with $\vphi(x,0^+) = \vphi(x,0^-)$. In our setup, the fields are defined on a ring of circumference $L$ in space, and imaginary time extends as $\tau \in (-\infty,\infty)$, so the Euclidean spacetime is an infinite cylinder of circumference $L$ (see the left cylinder in Fig.~\ref{setup}). As discussed previously, in this field-theory language, performing a projective measurement of the charge on the ground state via the operator $M_\m$ (see Eq.~\eqref{measurementop}) amounts to fixing the value of the field $\vphi$ in the measured region, i.e., imposing Dirichlet boundary conditions. Therefore, $\tr \rho_\m$ is given by the constrained path integral
\begin{equation}
\label{constrainedpathint}
    \tr \rho_\m = Z_\m
    = \int \mc D[\vphi]\, \exp[-S(\vphi)]\,
    \delta\bigl(\vphi(x,\tau=0) - \m(x)\bigr),
\end{equation}
with $\vphi(x,0^{-}) = \vphi(x,0^{+})$. Here $\m(x)$ denotes a putative continuum profile corresponding to the lattice measurement outcome $\m$ (although, as discussed previously, there is no simple, well-established lattice--continuum correspondence at the level of individual outcomes).
\begin{figure}[t!]
    \centering
    \includegraphics[width=0.7\linewidth]{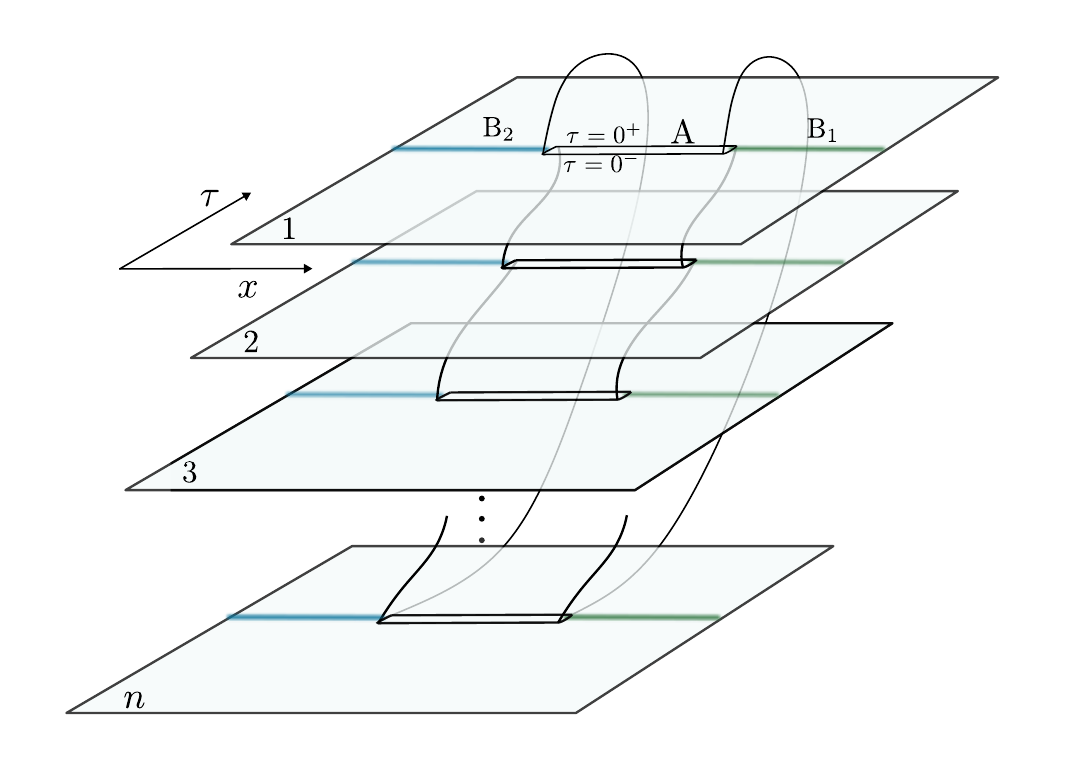}
    \caption{\textbf{$n$-sheeted Riemann surface.} Illustration of the gluing condition in region $A$ from Eq.~\eqref{gluing}, following Refs.~\cite{calabrese2009entanglement, Pasquale_Calabrese_2004}. For simplicity, the constituent manifolds are drawn as infinite planes rather than cylinders, and region $C$ is omitted. The measurement region $B_1$ ($B_2$) is depicted in green (blue). }
  \label{riemann}
\end{figure}
An analogous path-integral representation can be given for $\tr \rho_{\m,A}^n$~\cite{calabrese2009entanglement, Pasquale_Calabrese_2004}, which appears in Eq.~\eqref{avgmiereplica}, where $\rho_{\m,A} = \Tr_{C}\rho_\m$. To construct it, we first consider $n$ copies of the full density matrix $\rho$. In the path-integral language, this corresponds to $n$ copies of the geometry used to represent $\Tr \rho$, but with the fields left unconstrained at the imaginary time $\tau=0$ (i.e., without imposing periodic boundary conditions). On each copy we then impose the measurement constraint as in Eq.~\eqref{constrainedpathint}, thereby obtaining $\rho_\m^{\otimes n}$. The reduced density matrix $\rho_{\m,A}^{\otimes n}$ is obtained by tracing out region $C$, which at the level of the path integral amounts to gluing the fields at $\tau=0$ along $C$ within each replica. To implement the $n$-fold product and take the overall trace in $A$, we must further impose cyclic boundary conditions for $x \in A$ at $\tau = 0$:
\begin{equation}
\label{gluing}
\begin{aligned}
    \vphi^{(1)}(x,0^+) &= \vphi^{(2)}(x,0^-),\\
    \vphi^{(2)}(x,0^+) &= \vphi^{(3)}(x,0^-),\\
    &\;\vdots\\
    \vphi^{(n)}(x,0^+) &= \vphi^{(1)}(x,0^-).
\end{aligned}
\end{equation}

The resulting manifold is the $n$-sheeted Riemann cylinder, which we denote by $\mc M_n$ (see an illustration of the infinite plane version in Fig.~\ref{riemann}) \cite{calabrese2009entanglement, Pasquale_Calabrese_2004}. We can therefore write $\tr \rho_{\m,A}^n = Z_{\mc M_n,\m},$ where $Z_{\mc M_n,\m}$ is the $n$-sheeted analogue of $Z_\m \equiv Z_{\mc M_1,\m}$. With these path integral representations, we can re-cast the replica trick in a more intuitive form of a free energy difference:
\begin{align}
\label{replicapartition}
    \ov{S^n_{\m,A}} &= \frac{1}{1-n}\lim_{k\ra 0}\partial_k[\log \mc Z_A - \log\mc Z_0] \\
    &=\frac{1}{n-1}\lim_{k\ra 0}\partial_k[F_A -F_0], 
\end{align}
where we introduce the replica partition functions 
\begin{align}
    \label{miereplicapartitionfunctions1}
    \mc Z_A &= \ov{(\tr\rho^n_{\m,A})^k}=\ov{(Z_{\m,\mc M_n})^k},\\
    \label{miereplicapartitionfunctions2}
    \mc Z_0 &= \ov{\tr(\rho_{\m})^{nk}}=\ov{(Z_{\m})^{nk}},
\end{align}
and where $F_{A,0}=-\log\mc Z_{A,0}$. To obtain MIE, we must then evaluate the ``averaged'' replica partition functions $\mc Z_{A,0}$, which were evaluated analytically in \cite{khanna2025measurementinducedentanglementconformalfield} for all replica indices $(n,k)$. We here point out that $\mc Z_0$ is already an object of interest that characterizes, for example, the classical Shannon-entropy of the measurement probabilities (see Refs.~\cite{PhysRevB.84.195128, PhysRevB.90.045424}). It contains universal information about the distribution of the Born probabilities, a quantity of interest in several recent works~\cite{PhysRevX.14.041051, mcginley2025scroogeensemblemanybodyquantum} that study ensembles of states from chaotic unitary evolutions. $\mc Z_A$ on the other hand is necessary when discussing entanglement as seen above. However, the above replica trick as given above alone is not enough to generate the complete distribution or the moments of MIE. 
Mathematically, this is because, unlike in Eq.~\eqref{replicapartition}, one cannot fully factorize the dependence on $\tr \rho^n_{\m,A}$ and $\tr \rho_\m$ when considering moments of MIE. Instead, one must deal with \emph{joint} moments of the form $(\tr \rho^n_{\m,A})^{k_1}\tr \rho_{\m}^{k_2}$. To see this, we start from the cumulant expansion for the \textit{normalized} post-measurement density matrix $\tilde\rho_{\m,A }=\tr \rho^n_{\m,A}/(\tr\rho_\m)^n$ in terms of $S^n_{\m,A}$ as
\begin{equation}
\label{cumulantexpansion}
    \log\overline{\left[\frac{\tr\rho^n_{\m,A}}{(\tr\rho_\m)^n}\right]^k}
    = k(1-n)\,\overline{S^n_{\m,A}}
      + (1-n)^2\frac{k^2}{2}\,\overline{\bigl(S^n_{\m,A} - \overline{S^n_{\m,A}}\bigr)^2}
      + \dots.
\end{equation}
Motivated by this structure, we introduce a two-parameter \emph{generalized} replica partition function
\begin{equation}
\label{replicapartitionfunction}
    \mc Z_A(k_1,k_2)
    = \overline{(\tr\rho^n_{\m,A})^{k_1}\,\tr\rho^{k_2}_{\m}},
\end{equation}
in terms of which the $l$-th cumulant of $S^n_{\m,A}$ can be written as
\begin{equation}
\label{cumulantreplica}
    \kappa_l[S^n_{\m, A}]
    = \frac{1}{(1-n)^l}
      \lim_{k\to 0}
      \partial_k^{(l)} \log \mc Z_A(k_1,k_2)\big|_{k_1 = k,\; k_2 = -nk},
\end{equation}
where the derivatives are taken along the line $k_1 = k$, $k_2 = -nk$ and evaluated at $k\to 0$. Using the path-integral representation developed above, the generalized partition function can be written as
\begin{equation}
\label{finalgeneralizedpartition}
\mc Z_A(k_1,k_2)
= \overline{(Z_{\m,\mc M_n})^{k_1}\,(Z_{\m})^{k_2}},
\end{equation}
where the remainder of this section will discuss the computation of this object.\\
Before proceeding, we note that Refs.~\cite{PhysRevB.84.195128, PhysRevB.90.045424}, which study $\mc Z_0$ in the distinct context of Rényi-Shannon entropies, discuss a boundary phase transition occurring above a threshold replica number $Q_c(g)=2/g$, translated here into our notation. In our setup, the total number of replicas appearing in $\mc Z_A(k_1,k_2)$ is $Q = n k_1 + k_2 + 1$, which naturally raises the question of whether analogous boundary transitions can affect our calculation. However, we are always interested in the replica limit $k_1,-k_2/n=k\rightarrow 0$, or equivalently $Q\rightarrow 1$. Therefore, such a transition can only be encountered if $Q_c(g)\leq 1$, namely for $g\geq 2$, independently of the Rényi index $n$ or the cumulant order. By contrast, in this work we benchmark our results against the critical XXZ chain, for which $0<g<1$ throughout the critical phase $-1<\Delta<1$, using $\Delta=-\cos(\pi g)$. The boundary transition discussed in Refs.~\cite{PhysRevB.84.195128, PhysRevB.90.045424} is therefore outside the regime relevant to our analysis.
\subsection{Conformal Map and Quantum-Classical Split}
\label{conformalmapandqclassicalsplit}
Up to this point, our analysis applies to any field theory with a well-defined Euclidean action. Specializing to the free-boson theory \eqref{euclideanaction} allows for further simplification of the replica partition function \eqref{replicapartitionfunction} using the CFT toolkit. However, there is a caveat to being able to use CFT methods — it effectively assumes that \textit{all} measurement outcomes $\m$ flow to conformally invariant (Dirichlet) boundary conditions with a \textit{uniform} field value. There exist boundary conditions on the lattice, such as the ferromagnetic $\ket{\uparrow\uparrow\uparrow\dots}$, that in fact do not flow to such conformal boundary conditions under RG \cite{Stéphan_2014,PhysRevB.93.125139}, and are not even captured in the framework of bosonization which assumes a linear dispersion. Despite this, outcomes of this nature do not seem to play a role in the cumulants of MIE. A good reason for this is because such outcomes have global charge $\sim O(L)$ and are more than exponentially rare in their occurrence at half-filling. From a CFT point of view, such a breaking of conformal invariance is only true near the boundary and so one may expect contributions from such ``bad'' boundary conditions to be small in practice when considering an average over all boundary conditions. Previous work \cite{Lin2023probingsign, PhysRevB.109.195128, khanna2025measurementinducedentanglementconformalfield} shows numerical evidence for the MIE being conformally invariant suggesting that this is indeed true, thereby justifying the usage of standard CFT machinery in this case. In this work, we also find this to be true of the higher cumulants, hence also justifying this for the calculation of the cumulants presented in this work. Finally, for the reasons outlined above, we regard the use of the conformal map as justified. Importantly, however, we still keep the measurement outcome $\m(x)$ explicitly as a random, spatially varying Dirichlet boundary field; see, for example, Eq.~\eqref{bcprerotn}. Thus, we do not assume uniform boundary conditions at the level of field theory.\\
With this in mind, we continue with the simplification of \eqref{finalgeneralizedpartition} by performing a conformal map $\bar w_n(z)$ from the $n$-sheeted Riemann surface $\mc M_n$ with slits in $B$ to a cylinder $\mc C(n)$ of length $h(\zeta)/n$ with \cite{Rajabpour_2016, Kythe2019handbook, antonini_holographic_2022}
\begin{equation}
\label{hzeta}
    h(\zeta)  = 2\pi \frac{\mc K(k)}{\mc K(\sqrt{1-k^2})},
\end{equation}
and circumference $\beta = 2\pi$ (see bottom of Fig.~\ref{setup}), where $k = (1-\sqrt{1-\zeta})/(1+\sqrt{1-\zeta})$, and $\mc K(k) = \int_0^{\pi/2}d\theta/(\sqrt{1-k^2\sin^2\theta})$ is an elliptic integral. $\zeta$ here is the cross-ratio given by $\zeta = w_{12}w_{34}w^{-1}_{13}w^{-1}_{24}$ with $w_{ij} = (L/\pi)\sin(\pi(x_i-x_j)/L)$ being the chord length. Consequently, we will see that all ensuing expressions depend only on the conformally invariant variable $h(\zeta)$ generated by this map, making their conformal invariance manifest.

Such a map induces a change in the free-energy $F = -\log Z_{\mc M_n,\m}$ which factorizes into two parts \cite{Bimonte_2013, Rajabpour_2016}:
\begin{equation}
\label{freeenergysplit}
    F_{\mc M_n,\m} = F_{\mc C(n),\m} + F_n^{\mathrm{geom}},
\end{equation}
where the first term $F_{\mc C(n), \m} = -\log Z_{\mc M_n,\m}$ is the free energy of the compact free boson on the cylinder $\mc C(n)$ with boundaries (colored in blue and green in Fig.~\ref{setup}) encoding $\m$, and the latter is termed the ``geometric'' contribution (since it depends purely on the map $\bar{w}_n(z)$) and satisfies $\delta F_n^{\mathrm{geom}}/{\delta l} = \frac{ic}{12\pi}\oint_{C_2}\{\bar{w}_n(z),z\}$ \cite{Bimonte_2013, Rajabpour_2016}, with $c(=1)$ being the central charge. In our previous work \cite{khanna2025measurementinducedentanglementconformalfield}, we showed that the geometric free-energy $F^{\mathrm{geom}}_n$, although non-zero by itself, and containing non-universal contributions, vanishes when we perform the replica trick for the MIE due to the relation $F^{\mathrm{geom}}_n=nF^{\mathrm{geom}}_1$. This cancellation, however, is specific to the cylinder mapping $\mc C(n)$ and need not persist on other target manifolds. For example, mapping to an annulus $\mc A(n)$ with inner and outer radii—as in Refs.~\cite{Rajabpour_2016, najafi_entanglement_2016, antonini_holographic_2023}—yields an analogous decomposition to \eqref{freeenergysplit}. In this geometry, however, the geometric term does not cancel in the replica trick since $F^{\mathrm{geom}}_{\mc A(n)} \neq n\,F^{\mathrm{geom}}_{\mc A(1)}$. Consequently, the geometric piece must be retained explicitly and is evaluated via the contour integral mentioned above. Regardless of the target manifold, note that the resulting decomposition must reproduce the same full free energy $F_{\mc M_n,\m}$. For our purposes, the cylinder map $\mc C(n)$ is therefore preferable—it reduces the problem to standard cylinder partition functions and avoids the additional contour–integral subtleties of the annular case, allowing us to write down closed form expressions unlike in Refs.~\cite{Rajabpour_2016, najafi_entanglement_2016, antonini_holographic_2023}.

Through a similar line of reasoning, the geometric part when mapping to $\mc C(n)$ does not contribute to the cumulants of MIE. We therefore drop the geometric factor in $Z_{\mc M_n,\m}$ altogether and write 
\begin{equation}
\label{zaoncn}
\mc Z_A \sim \ov{ (Z_{\mc C(n),\m})^{k_1}(Z_{\mc C(1), \m})^{k_2}}=\sum_\m (Z_{\mc C(1),\m})^{k_2+1}(Z_{\mc C(n),\m})^{k_1} ,
\end{equation}
where we have replaced the Riemann cylinder $\mc M_n$ with the finite cylinder $\mc C(n)$ and used $p_\m = \tr\rho_\m\sim Z_{\mc C(1),\m}$. The partition function $Z_{\mc C(n),\m}$ on the cylinder $\mc C(n)$ can be further simplified by a ``quantum-classical'' splitting of the bosonic field as $\vphi=\vphi_{\rm cl,\m}+\vphi_q$ which causes a corresponding split in the partition function
\begin{equation}
\label{qclassicalsplit}
    Z_{\mc C(n),\m} = Z^q_{\mc C(n), D}Z^{\rm cl}_{\mc C(n), \m}=Z^q_{\mc C(n), D}\sum_w\exp\left[-S_{\mc C(n)}[\vphi_{\rm cl,\m}]\right]. 
\end{equation}
The first factor is the ``quantum fluctuations’’ contribution, obtained by summing
over bosonic oscillator modes on the cylinder with Dirichlet boundary conditions 
($\vphi|_B=0$, denoted by the subscript $D$):
\begin{equation}
\label{qfluctuating}
    Z^q_{\mc C(n), D}
    = e^{-\beta E_0}
\prod_{s=1}^{\infty}\Biggl[\sum_{n=0}^{\infty} e^{-n\beta\omega_s}\Biggr]
    = \frac{1}{\eta(q_n)},
\end{equation}
where the allowed frequencies on $\mc C(n)$ are $\omega_s = s\pi/(h(\zeta)/n)$, and \(E_0=\frac{1}{2}\sum_{s=1}^{\infty}\omega_s \allowbreak = \frac{-\pi}{24 (h(\zeta)/n)}\) is the ground state energy. In CFT parlance, this is just the Dedekind eta function $1/\eta(q_n)$ with $q_n = e^{-\pi \beta /(h(\zeta)/n)}$. The second factor in \eqref{qclassicalsplit} is the partition function for the classical field $\vphi_{\rm cl,\m}$ satisfying the Laplace equation $\nabla^2\vphi=0$ on $\mc C(n)$ with measurement boundary conditions $\vphi_{\rm cl,\m}|_{C_1} = \m_1(\theta) + 2\pi w$ and $\vphi_{\rm cl,\m}|_{C_2} = \m_2(\theta)$, where $w\in\mathbb{Z}$ is the winding number arising from the identification $\vphi\sim\vphi+2\pi$\footnote{It suffices to allow winding on only one boundary
(here $C_1$): a constant shift of $\vphi$ can translate the winding between $C_1$
and $C_2$.}, and where we split $\m=\m_1\cup\m_2$ across $C=C_1\cup C_2$. Here $\m_{1,2}(\theta)$ are coarse-grained versions of the lattice measurement outcomes along the angular coordinate $\theta$ of the cylinder. Note that we do not restrict $\m(\theta)$ to be conformally invariant and will ultimately sum over all (random) profiles $\m(\theta)$.

\subsection{The Classical Winding Contribution}
In this section we focus on the classical contribution $Z^{\rm cl}_{\mc C(n),\m}=\sum_w\exp\left[-S_{\mc C(n)}[\vphi_{\rm cl,\m}]\right]$ in \eqref{qclassicalsplit} to the replica partition function \eqref{finalgeneralizedpartition}. Since the quantum fluctuating contribution \eqref{qfluctuating} is measurement-independent and thereby factors out of the average in \eqref{finalgeneralizedpartition}, it is indeed the only non-trivial measurement-dependent part that needs careful evaluation. So far, from Eqs.~\eqref{zaoncn}, \eqref{qclassicalsplit}, and~\eqref{qfluctuating}, we have
\begin{equation}
\label{replicapartitionsimplified1}
    \mc Z_A(k_1,k_2)\sim \frac{1}{\eta(q_n)^{k_1}}\frac{1}{\eta(q_1)^{k_2+1}}\sum_{\vec{w}\in\mathbb{Z}^{k_1+k_2}}\sum_\m \exp\left[-\sum_{i=1}^{k_1}S_{\mc C(n)}\left[\vphi^{(i)}_{\rm cl,\m}\right]-\sum_{i=k_1+1}^{k_2+k_1+1}S_{\mc C(1)}\left[\vphi^{(j)}_{\rm cl,\m}\right]\right], 
\end{equation}
where we have dropped the geometric factor discussed previously, and we have used the fact that $p_\m=\tr\rho_\m$ resulting in an additional replica on $\mc C(1)$. We however have one less winding number in the tuple $\vec{w} = (w_1,\dots,w_{k_1}, w_{k_1+1}\dots,w_{k_2+k_1})$ since we consider winding numbers relative to the last replica. One can justify the consideration of relative winding at a more technical level \cite{khanna2025measurementinducedentanglementconformalfield} but intuitively, it is easy to see that only relative winding numbers have any physical meaning. For example, one can increase all the winding numbers uniformly and no observable change can be detected. 

The non-trivial interaction among the above replicas is encoded in the common boundary condition they obey at the boundaries. As a first step in simplifying the above, it is helpful to recast every term to be on the same manifold, say $\mc C(n)$. We can do this by noting that
\begin{equation}
\label{freebosonactionclassical}
        S_{\mc C(1)}[\vphi_{{\rm cl},\m}] = \frac{g}{4 \pi} \int_0^{\beta = 2 \pi} d \tau \int_0^{h(\zeta)} dy \left(\frac{\delta\m }{ h(\zeta)}\right)^2 = \frac{g}{2}\frac{(\delta \m)^2}{h(\zeta)} = \frac{g}{2}\frac{(\delta \m/\sqrt{n})^2}{h(\zeta)/n} = S_{\mc C(n)}[\tilde{\vphi}_{{\rm cl},\m}], 
\end{equation}
where $\tilde \vphi_{\rm cl,\m}=\vphi_{\rm cl,\m}/\sqrt{n}$. Rewriting $S_{\mc C(1)}[\vphi]$ in \eqref{replicapartitionsimplified1} using the above identity, we effectively trade
a mismatch in the domains of the replicas to a difference in boundary conditions between them, namely
\begin{equation}
\label{bcprerotn}
\begin{split}
        \vphi^{(i)}_{{\rm cl},\m}\vert_{B_1} &= {\m_1}(\theta)+ 2\pi w_i\;\;\;\;i={1,\dots ,k_1,}\\
    \tilde\vphi^{(j)}_{{\rm cl},\m}\vert_{B_1} &=\frac{\m_1(\theta)}{\sqrt{n}}+ \frac{2\pi w_i}{\sqrt{n}}\;\;\;\;i={k_1+1,\dots ,k_1+k_2+1},
\end{split}
\end{equation}
with identical boundary conditions on $B_2$ without any winding, and where $w_{k_2+k_1+1}=0$ with the other windings taken relative to it. At this stage it is useful to perform a basis transformation such that in this new basis, only \textit{one} copy of the field has boundary conditions that depend on the measurement outcomes $\m$ and the rest are only winding number dependent. In this case it is simply the transformation that rotates the vector $\vec{\mu} = (1,\dots,1,1/\sqrt{n},\dots ,1/\sqrt{n})$ to $\vec{\nu} =||\vec{\mu}|| (0,\dots,1)$ and is hence the reflection matrix
\begin{equation}
\label{basisrotn}
    \mc R_{k_1+k_2+1} = \mathbb{I} - 2\frac{\vec{\gamma}\vec{\gamma}^T}{\langle\vec{\gamma}, \vec{\gamma}\rangle},
\end{equation}
where $\vec{\gamma} = \vec{\mu}-\vec{\nu}$. Such a trick of rotating fields to ``cancel" measurement dependence was first introduced by Fradkin and Moore in a very different context \cite{PhysRevLett.97.050404}. The authors however neglected the winding contributions that were later reinstated in future works \cite{PhysRevB.79.115421, oshikawa2010boundary, PhysRevLett.107.020402, Zhou_2016}. As we shall see shortly, in our case it is indeed these winding contributions that give rise to novel critical exponents and qualitative behavior for the cumulants of MIE. Explicitly, after transforming from the fields $\bm{\vphi} = (\vphi^{(1)},\dots,\vphi^{(k_1)},\tilde{\vphi}^{(k_1+1)},\dots,\tilde{\vphi}^{(k_1+k_2+1)} )$ to the new basis $\bm{\bar{\vphi}}=(\bar{\vphi}^{(1)},\dots,\bar{\vphi}^{(k_1)},\bar{\vphi}^{(k_1+1)},\dots \allowbreak ,\bar{\vphi}^{(k_1+k_2+1)} )$, the boundary conditions in Eq.~\eqref{bcprerotn} become
\begin{equation}
\label{bcpostrotn}
\begin{split}
    \bar{\vphi}^{(i)}_{\rm cl,\m}|_{B_1} &= 2\pi [M_{k_1+k_2}\Lambda]_{ij}w_j\;\;\;\;i={1,\dots ,k_1+k_2}\\\\
    \bar{\vphi}^{(k_1+k_2+1)}_{\rm cl,\m}|_{B_1} & = \sqrt{\frac{nk_1+k_2+1}{n}} \m_1(\theta) + 2\pi \sqrt{\frac{1}{nk_1+k_2+1}}\left(\sum_{i=1}^{k_1}\sqrt n w_i+\sum_{i=k_1+1}^{k_2+k_1} \frac{w_i}{\sqrt n}\right),
\end{split}
\end{equation}
where $M_{k_1+k_2}$ is the top left $(k_1+k_2)\times(k_1+k_2)$ block of $\mc R_{k_1+k_2+1}$ and the matrix $\Lambda = \mathrm{diag}(1\dots\allowbreak,1,1/\sqrt{n},\allowbreak\dots,1/\sqrt{n})$ encodes the $n$-dependence of the winding term in Eq.~\eqref{bcprerotn}. By construction, we see that the measurement dependence resides only in the boundary condition of the last replica, while the other replicas contribute solely through windings that account for compactification in the rotated basis. Since the combination within the measurement summation in Eq.~\eqref{replicapartitionsimplified1} remains invariant under the rotation, we may then factor out all but the $(k_1+k_2+1)$th replica and define a winding contribution
\begin{equation}
\label{windinggeneralized}
    \mc W^{(n)}_{k_1,k_2}=\sum_{\vec{w}\in\mathbb{Z}^{k_1+k_2}}\exp\left[-\sum_{i=1}^{k_1+k_2}S_{\mc C(n)}\left[\bar{\vphi}^{(i)}_{\rm cl}\right]\right],
\end{equation}
involving $k_1+k_2$ replicas. The summation $\sum_\m e^{-S[\bar{\vphi}^{(k_1+k_2+1)}_{\rm cl,\m}]}$ over the final replica simplifies using the relation $\sum_\m Z_{\m, \mc C(n)}=\frac{1}{\eta(q_1)}\sum_\m e^{-S[\bar{\vphi}^{(k_1+k_2+1)}_{\rm cl,\m}]} \equiv 1$ and drops out of the replica trick since it is $k_1, k_2$-independent, leaving only the winding piece. With this we have
\begin{equation}
\label{generlizedreplicapartition}
    \mc Z_A(k_1,k_2) \sim \frac{1}{\eta(q_n)^{k_1}}\frac{1}{\eta(q_1)^{k_2+1}} \mc W^{(n)}_{k_1,k_2},
\end{equation}
where we have dropped terms that are inessential for the replica limit of Eq.~\eqref{cumulantreplica}. Thus, the net effect of summing over boundary conditions $\m$ across replicas is encoded in the winding function above, which arises from the change of basis \eqref{basisrotn}. 
The winding function is easily given by directly evaluating the classical action in \eqref{windinggeneralized} where the boundary conditions for the fields are now given in Eq.~\eqref{bcpostrotn}. This results in 
\begin{equation}
\label{generalizedwindingqn}
    \mc W^{(n)}_{k_1,k_2} = \sum_{\vec{w}\in\mathbb{Z}^{k_1+k_2}}\exp\left[-\frac{gn}{2h(\zeta)}(2\pi)^2\vec{w}^T\Lambda^T M^T_{k_1+k_2}M_{k_1+k_2}\Lambda \vec{w}\right] = \sum_{\vec{w}\in\mathbb Z^{k_1+k_2}}q_n^{g\vec{w}^T T_{k_1+k_2} \vec{w}},
\end{equation}
where we have defined $T_{k_1+k_2} = \Lambda^T M_{k_1+k_2}^TM_{k_1+k_2}\Lambda$ and $q_n= e^{-\pi n\beta/h}=e^{-2\pi^2n/h(\zeta)}$ was introduced in Eq.~\eqref{qfluctuating}. The obtained structure of the winding function is intuitive: it resembles the classical contribution to the usual free-boson partition function on $\mc C(n)$, $Z_{\mc C(n),\delta\vphi=0}\sim\allowbreak \sum_w q^{gw^2}$ with the conformal boundary condition $\delta_{\vphi}=0$, except that now its $(k_1+k_2)-$times replicated and the replicas are coupled through $T_{k_1+k_2}$. The above summation must now be analytically continued in order to enable the replica limit in \eqref{cumulantreplica}. We defer the details of analytic continuation of the above to the Appendix \ref{analyticcontwn} and give the final result:
\begin{equation}
\label{generalizedwinding}
    \mc W_{k_1,k_2}^{(n)}= \sqrt{\frac{(nk_1+k_2+1)g}{2 \pi h}}\int_{-\infty}^{\infty} d\delta_\vphi e^{- \frac{g\delta_{\vphi}^2}{2h}}\left[\sum_{w\in\mathbb Z}q_1^{g(w+\delta_\vphi/2\pi)^2}\right]^{k_2} \left[\sum_{w\in\mathbb Z}q_n^{g(w+\delta_\vphi/2\pi)^2}\right]^{k_1}, 
\end{equation}
where $\delta_\vphi=\vphi_{\rm cl}|_{B_1} - \vphi_{\rm cl}|_{B_2}$ labels all \emph{conformal} boundary conditions of the free-boson theory. Given that we  constructed the conformal map disregarding potential non-conformal boundary terms, the emergence of only conformal boundary conditions above is somewhat natural. What is however nontrivial is that, in computing $\mc W_{k_1,k_2}^{(n)}$, we still sum over \textit{all} $\m$, which a priori includes non-conformal outcomes. Nevertheless, the full sum remains conformally invariant and collapses to an explicit integral over conformal boundary conditions, indicating that non-conformal contributions effectively drop out, and that most outcomes flow to conformally invariant boundary conditions under RG.
\section{Results}
\subsection{``Born-Averaging'' over Dirichlet BCs and Cumulants of MIE}
\label{bornaveragingandcumulants}
At first glance, the expressions leading to Eqs.~\eqref{generlizedreplicapartition} and~\eqref{generalizedwinding} may appear opaque. However, one can re-write them in a more suggestive form
\begin{equation}
\label{zaprobability}  
\mc Z_A(k_1,k_2) = \ov{(\tr\rho^n_{\m,A})^{k_1}\tr\rho^{k_2}_{\m}} \sim  \int_0^{2\pi }d\delta_\vphi\; Z_{\mc C(1), \delta_\vphi} (Z_{\mc C(n), \delta_{\vphi}})^{k_1} (Z_{\mc C_1, \delta_{\vphi}})^{k_2},
\end{equation}
where we have suppressed an overall normalization. This representation of the replica partition function $\mc Z_A(k_1,k_2)$ (which is defined as a Born average of the mixed moment $(\tr \rho^n_{\m,A})^{k_1}\tr \rho^{k_2}_{\m}$) admits a natural physical interpretation: Born averaging over measurement outcomes can be viewed at low energies as averaging over\textit{ conformal boundary conditions}, with each outcome $\delta_{\vphi}$ weighted by its ``Born probability'' $\sim Z_{\mc C(1), \delta_{\vphi}}$, as one would intuitively expect from a stat-mech notion of averaging. Furthermore, $\tr\rho^n_{\m,A}$ and $\tr\rho_{\m}$ too are replaced by their respective low-energy partition function representations $Z_{\mc C(n),\delta_{\vphi}}$ and $Z_{\mc C(1),\delta_{\vphi}}$ with boundary condition labeled via $\delta_{\vphi}$. Such an interpretation was first noted in our previous work \cite{khanna2025measurementinducedentanglementconformalfield}, where we further noted that the MIE could be written, in similar spirit as above, as an average of \textit{forced} MIE (denoted $\mief(\delta_{\vphi})$) over the boundary conditions labeled by $\delta_{\vphi}$:
\begin{equation}
\label{averagemieprobinterp}
\ov{S^{(n)}_{\m,A}} = \int_{0}^{2\pi} d\delta_{\vphi}\; p(\delta_{\vphi}) \mathrm{MIE^{(n)}_F}(\delta_{\vphi}), 
\end{equation}
where $\mief$, which is the MIE when one fixes a specific outcome (labeled here by $\delta_\vphi$ in the continuum limit) takes the form \cite{PhysRevB.92.075108, Rajabpour_2016, khanna2025measurementinducedentanglementconformalfield}
\begin{equation}
\label{mief}
\mathrm{MIE^{(n)}_F}(\delta_{\vphi}) = \frac{1}{1-n}\log\frac{Z_{\mc C(n), \delta_{\vphi}}}{Z^n_{\mc C(1), \delta_{\vphi}}},
\end{equation}
with $Z_{\mc C(n), \delta_{\vphi}} = \eta(q_n)^{-1}\sum_{w\in\mathbb Z} q_n^{g(w+\frac{\delta_{\vphi}}{2\pi})^2}$, and 
\begin{equation}
\label{pdeltaphi}
    p(\delta_{\vphi}) = \sqrt{\frac{g}{2\pi h}}\sum_{l\in\mathbb Z}q_1^{2g(l+\frac{\delta_{\vphi}}{2\pi})^2} = \frac{Z_{\mc C(1), \delta_{\vphi}}}{\int_0^{2\pi} d\delta_{\vphi}'\;Z_{\mc C(1),\delta_{\vphi}'}}
\end{equation}
acts as a probability distribution over $\delta_{\vphi}$. In this work, we confirm this interpretation by using it to evaluate \textit{all} higher cumulants of MIE. In particular, we emphasize that $p(\delta_{\vphi})\propto Z_{\mc C(1),\delta_{\vphi}}$ serves as a bonafide probability distribution enabling a systematic computation of higher cumulants of MIE by averaging over moments of the forced MIE with the distribution $p(\delta_{\vphi})$. For instance, one can already write the second cumulant of MIE by taking $p(\delta_{\vphi})$ as the distribution, in a similar way as the mean \eqref{averagemieprobinterp}, giving
\begin{equation}
\label{secondcumulantguess}
\kappa_2[S^n_{\m,A}] = \ov{(S^{n}_{\m,A})^2}-\ov{S^{n}_{\m,A}}^2
= \int_{0}^{2\pi} d\delta_{\vphi}\; p(\delta_{\vphi}) (\mathrm{MIE^{(n)}_F}(\delta_{\vphi}))^2
- \left(\int_{0}^{2\pi} d\delta_{\vphi}\; p(\delta_{\vphi}) \mathrm{MIE^{(n)}_F}(\delta_{\vphi})\right)^2.
\end{equation}
Indeed, we find that taking the replica limit in \eqref{cumulantreplica} with the derived generalized replica partition function \eqref{generlizedreplicapartition} in the previous section reproduces precisely \eqref{secondcumulantguess} for the second cumulant. The connection with the forced MIE \eqref{mief} that appears in the expression for MIE (Eq. \eqref{averagemieprobinterp}) can more generally be shown for higher cumulants by first re-writing
\begin{equation}
\left(\frac{Z_{\mc C(n), \delta_{\vphi}}}{Z^n_{\mc C(1),\delta_{\vphi}}}\right)^k = \exp\left[k\log\left(\frac{Z_{\mc C(n), \delta_{\vphi}}}{Z^n_{\mc C(1),\delta_{\vphi}}}\right)\right] = \exp\left[k(1-n)\mathrm{MIE^{(n)}_F}(\delta_{\vphi})\right],
\end{equation}
which lets us recast the expression \eqref{cumulantreplica} for cumulants of MIE (using \eqref{zaprobability}) as
\begin{equation}
\label{cumulantfinalexpression}
\boxed{
\begin{gathered}
  \kappa_{l}\bigl[S^{(n)}_{\m,A}\bigr]
  = \frac{1}{(1-n)^l}
    \lim_{k\to 0}
    \partial^{(l)}_{k}
    \log \mathbb{E}_{\delta_{\vphi}\sim p(\delta_{\vphi})}
    \bigl[e^{k(1-n)\,\mathrm{MIE}^{(n)}_{F}(\delta_{\vphi})}\bigr],
  \\
  \mathbb{E}_{\delta_{\vphi}}[\cdot]
  = \int d\delta_{\vphi}\, p(\delta_{\vphi})[\cdot],
\end{gathered}
}
\end{equation}
with $p(\delta_{\vphi})$ defined in Eq.~\eqref{pdeltaphi}, and where $\log\mathbb{E}_{\delta_{\vphi}}\left[e^{{k(1-n)\mathrm{MIE^{(n)}_F})}}\right]$ acts as a cumulant generating function for MIE, given in terms of $\mief$. The above is a central result of our work. It shows that MIE and its higher cumulants for 1d quantum critical states governed by the compact free-boson theory can be expressed as cumulants of the random variable $\mathrm{MIE^{(n)}_F}(\delta_{\vphi})$ with respect to the probability distribution $p(\delta_{\vphi})$, where $\delta_{\vphi}$ labels the measurement outcomes. Finally, we benchmark our predictions on the XXZ chain \cite{Peschel_2009}. We find very good agreement across a range of Rényi indices for the second and third cumulants (see Fig.~\ref{plots}a, ~\ref{plots}b, ~\ref{plots2}a, and~\ref{plots2}b). The numerical procedure is described in Appendix~\ref{numericsappendix}.

\subsection{Universal Scaling of Cumulants in the $\zeta\ra0 $ Limit}
\label{universalscalingcumulantszeta0}
The cumulants of MIE are most revealing of measurement physics in the $\zeta \to 0$ limit, corresponding to maximal separation between the unmeasured parties. This regime is especially relevant for two reasons. First, MIE was originally proposed as a measure of average “localizable entanglement’’ (LE) \cite{PhysRevLett.92.027901, PhysRevA.71.042306} between two \textit{distant} parties after measuring the rest of the system. Second, in this limit the non-trivial winding function \eqref{generalizedwinding}, absent in the forced case, governs critical behavior, revealing true measurement averaged contributions \cite{Lin2023probingsign, PhysRevB.109.195128, khanna2025measurementinducedentanglementconformalfield}. In particular, we find that the cumulants scale as, 
\begin{figure*}[t]
    \centering
    \includegraphics[width=\linewidth]{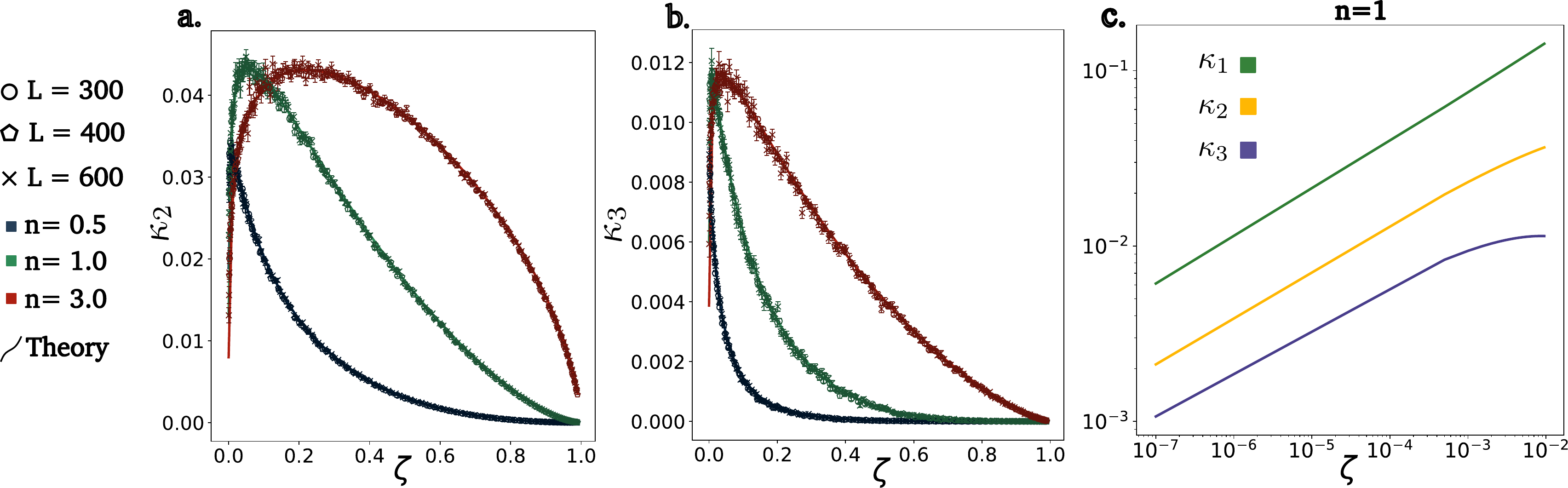}
    \caption{\textbf{Cumulants of MIE versus cross ratio for anisotropy $\Delta = 0$.} Markers show numerical results for cumulants of MIE for the XX chain for the setup in Fig~\ref{Fig1}, and solid curves are theoretical predictions from Eq.~\eqref{cumulantfinalexpression}. 
    (a) $\kappa_2 = \ov{S_{\m,A}^2} - \ov{S}_{\m,A}^2$. (b) $\kappa_3 = \ov{S_{\m,A}^3} - 3\,\ov{S_{\m,A}^2}\,\ov{S}_{\m,A} + 2\,\ov{S}_{\m,A}^3$. (c) Theoretical prediction extended to regimes of $\zeta$ where we see the scaling Eq.~\eqref{scalingcumulants} for $l=1,2,$ and $3$.}
    \label{plots}
\end{figure*}
\begin{equation}
\label{scalingcumulants}
\kappa_{l}\!\left[S^{(n)}_{\m,A}\right]
\underset{\zeta \to 0}{\sim}
\begin{cases}
\displaystyle \dfrac{\zeta^{g/2}}{\sqrt{\log(1/\zeta)}} & n > \tfrac{1}{2l}, \\[1.1ex]
\displaystyle \zeta^{g/2} & n = \tfrac{1}{2l}, \\[0.8ex]
\displaystyle \zeta^{2gnl(1-nl)} & 0 < n < \tfrac{1}{2l},
\end{cases}
\end{equation}
for $l\in\mathbb Z_{>0}$. The derivation is given in Appendix~\ref{cumulantsappendix}, where we evaluate the corresponding moments in the $\zeta\to0$ limit, allowing the moment index to be analytically continued away from integer values. For the integer cumulants considered here, the leading scaling is the same as that of the $l$-th moment: $\kappa_l[S^{(n)}_{\m,A}]=\overline{(S^{(n)}_{\m,A})^l}+$  products of lower moments, and these products are subleading as $\zeta\to0$. As seen above, we find a change in qualitative behavior of the $l$-th cumulant at $nl=1/2$. Notably, for large enough $n$, we see that \textit{all} cumulants of MIE scale in the same manner: $\zeta^{g/2}$ with a multiplicative $1/\sqrt{\log(1/\zeta)}$ factor which breaks the usual power law scaling that one typically encounters in CFTs. Using the formulas derived in the previous section, we confirm the identical scaling across cumulants explicitly at $n=1$ for $l=1,2,$ and $3$ (see Fig.~\ref{plots}c). We remark that we have used the analytic expressions for this since the small-$\zeta$ regime where these scalings coincide lies beyond the numerical window accessible in this work. 

Next, while the detailed derivation of the above scaling forms are in Appendix~\ref{cumulantsappendix}, we briefly explain the origin of the scaling forms above, especially of the rather odd (yet characteristic) $\sqrt{1/\log(1/\zeta)}$ factor. For this we consider the simpler case of MIE $(\kappa_{l=1})$ (Eqs.~\eqref{averagemieprobinterp} and~\eqref{mief}) since all cumulants scale in the same manner anyway. In the small-$\zeta$ regime, where most of the system is measured, the winding sum $\sum_{w\in\mathbb Z} q_n^{g(w+\frac{\delta_{\vphi}}{2\pi})^2}$ in the partition function $Z_{\mc C(n),\delta_{\vphi}}$ of the forced MIE \eqref{mief} furnishes the leading contribution. Collecting the leading winding terms for MIE and simplifying, we get that the leading behavior of MIE is controlled by the integral (see Appendix~\ref{cumulantsappendix})
\begin{equation}
\label{asymptoticint}
        \ov{(S^{(n)}_{\m, A})}\underset{\zeta \to 0}{\sim}    \sqrt{\frac{1}{h}}
      \int_{0}^{\pi}\! d\delta_{\varphi}\;
      e^{-\frac{g\,\delta_{\varphi}^2}{2h}}\left[\frac{
        e^{-\frac{2gn\pi^2}{h}\!\left(1-\frac{\delta_{\varphi}}{\pi}\right)}-n
                  e^{-\frac{2g\pi^2}{h}\!\left(1-\frac{\delta_{\varphi}}{\pi}\right)}}{1-n}\right],
\end{equation}
where the small $\zeta$ behavior is dictated by the quantity $h(\zeta)\underset{\zeta \to 0}{\sim} \pi^2/\log(1/\zeta)+\dots$ which also vanishes in this limit, and where we have expanded $q_n = e^{-2\pi^2n/h}$. While the terms in the square brackets are from the leading contributions of winding, the sharply peaked Gaussian $e^{-g\delta_\vphi^2/2h}$ is essentially the distribution $p(\delta_{\vphi})$ ``unwrapped'' out on the entire real line. If we momentarily ignore winding and retain only this Gaussian with its proper normalization $\sim 1/\sqrt{h}$, we have
\begin{equation}
\label{naiveint}
    \ov{(S^{(n)}_{\m, A})}\underset{h(\zeta) \to 0}{\sim}  \frac{1}{\sqrt{h}}\int_0^{\pi}d\delta_{\vphi}\;e^{-g\delta_\vphi^2/2h}\underset{h(\zeta) \to 0}{\sim} \zeta^{g/2},
\end{equation}
which already reproduces the $\zeta^{g/2}$ factor in \eqref{scalingcumulants}. Including the winding factor, which contributes as an exponential that is linear in $\vphi$, biases the Gaussian toward a nonzero mean boundary condition $\vphi\neq 0$. For example, taking the first term in the square brackets of \eqref{asymptoticint} yields
\begin{align}
     \ov{(S^{(n)}_{\m, A})}&\underset{h(\zeta)\to 0}{\sim}\frac{1}{\sqrt{h}}\int_0^{\pi}d\delta_{\vphi}\;e^{-\frac{g\vphi^2}{2h}}e^{-\frac{2\pi^2ng}{h}\left(1-\frac{\vphi}{\pi}\right)} \nonumber\\
     \label{cumulantexplanationint}
     &=\frac{e^{-\frac{2\pi^2gn(1-n)}{h}}}{\sqrt{h}}\int_{-2\pi n}^{-\pi(2n-1)}dt\; e^{-\frac{g}{2h}t^2}.
\end{align}
This immediately clarifies the transition at $nl=1/2$ (with $l=1$ here). For $n>1/2$, both integration limits lie on the \emph{same} side of the sharp peak at $t=0$, so the Gaussian contribution around the peak is excluded; relative to \eqref{naiveint}, the integral is suppressed by an additional factor $\sqrt{h}\sim 1/\sqrt{\log(1/\zeta)}$. In contrast, in \eqref{naiveint} the interval includes the peak, and no such suppression occurs. For $0<n<1/2$, the limits lie on either side of $t=0$ in \eqref{cumulantexplanationint}, so the peak region contributes and the extra $1/\sqrt{\log(1/\zeta)}$ factor is absent; the leading scaling is then set by the pre-factor $e^{-\frac{2\pi^2gn(1-n)}{h}}\underset{h(\zeta)\to 0}{\sim}\zeta^{2gn(1-n)}$, as stated in \eqref{scalingcumulants}. This analysis can be analogously carried out for $l>1$ and is given in the Appendix~\ref{cumulantsappendix}. \\ 
To end this section, we discuss the \textit{typical} value of MIE,
\begin{equation}
    \mathrm{MIE^{(n)}}_{\rm typical}(A) = e^{\mathbb E[\log S^{(n)}_{\m,A}]},
\end{equation}
which suppresses any rare large entanglement contributions to MIE. Having established\newline $\{\mathrm{MIE}_F(\delta_\varphi),p(\delta_\varphi)\}$  as the relevant ensemble for all cumulants of MIE, we can directly write
\begin{equation}
    \overline{\log S_\m^{(n)}} = \int d\delta_\varphi \; p(\delta_\varphi)\; \log\mathrm{MIE_F^{n}(\delta_\varphi)}. 
\end{equation} 
The $\zeta \ra 0 $ limit of this is controlled by the integral 
\begin{equation}
    \label{asymptoticintlog}
         \overline{\log S_\m^{(n)}}\underset{\zeta \to 0}{\sim}    \sqrt{\frac{1}{h}}
      \int_{0}^{\pi}\! d\delta_{\varphi}\;
      e^{-\frac{g\,\delta_{\varphi}^2}{2h}}\log\left[\frac{
        e^{-\frac{2gn\pi^2}{h}\!\left(1-\frac{\delta_{\varphi}}{\pi}\right)}-n
                  e^{-\frac{2g\pi^2}{h}\!\left(1-\frac{\delta_{\varphi}}{\pi}\right)}}{1-n}\right],
\end{equation}
analogous to Eq.~\eqref{asymptoticint}. Focusing on $n>1$, the logarithm is dominated by the linear term in $n$ in the numerator. Due to the sharply peaked gaussian of width $\mathcal O(\sqrt{h})$, the factor of $(1-\delta_\varphi/\pi) \approx 1 - \mathcal O(\sqrt{h})$, and the above integral scales as
\begin{equation}
     \overline{\log S_\m^{(n>1)}}\underset{\zeta \to 0}{\sim}\frac{1}{\sqrt h}\frac{-2\pi^2g}{h}\int_0^{\infty} d\delta_\varphi e^{-\frac{g\delta_\varphi^2}{2h}}  \underset{\zeta \to 0}{\sim} -2\pi^2g/h,  
\end{equation}
which implies 
\begin{equation}
\label{mietypical}
     \mathrm{MIE^{(n>1)}}_{\rm typical}(A) = e^{\mathbb E[\log S^{(n)}_{\m,A}]} \underset{\zeta \to 0}{\sim} \zeta^{2g}.
\end{equation}
First, note that the typical MIE scales quite differently than the averaged MIE (See Eq.~\eqref{scalingcumulants}) indicating that the distribution of post-measurement entanglement is likely not a simple one. We will explore this fact more in Sec.~\ref{universaldistmie} and show that this is indeed the case. Secondly, the exponent $2g$ of the typical MIE matches that of $\mathrm{MIE}_F(\delta_\varphi = 0)$ defined in Eq.~\eqref{mief} (see also Refs.~\cite{Rajabpour_2016, khanna2025measurementinducedentanglementconformalfield}). An example of such an outcome on the lattice is the Néel state $\ket{\uparrow\downarrow\uparrow\dots\uparrow\downarrow}$. 

 \begin{figure*}[t]
    \centering
    \includegraphics[width=\textwidth]{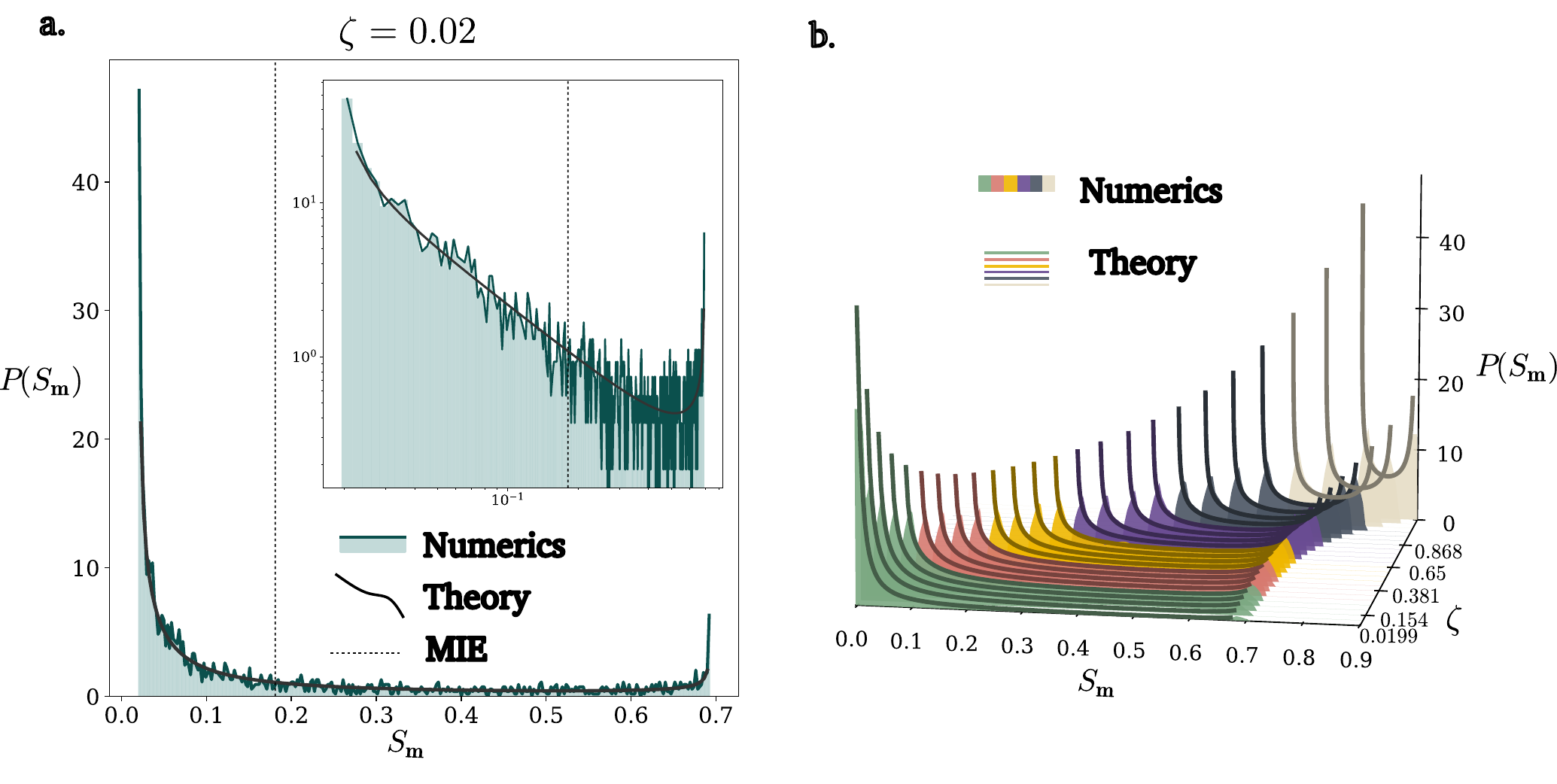}
    \caption{\textbf{Distribution of MIE.} \textbf{Left:} Comparison of the theory from Eq.~\eqref{miedist} and the numerical distribution $P(S_\m)$ for the XX chain at a fixed value of the cross-ratio $\zeta = 0.02$ and $L=600$ (see Fig.~\ref{Fig1} for the setup) with a log-log inset of the same. \textbf{Right:} Theory vs numerics comparison of the evolution of $P(S_\m)$ with cross-ratio $\zeta$.} 
    \label{miedistplots}
\end{figure*}
\begin{figure*}[t]
    \centering
    \includegraphics[width=0.9\linewidth]{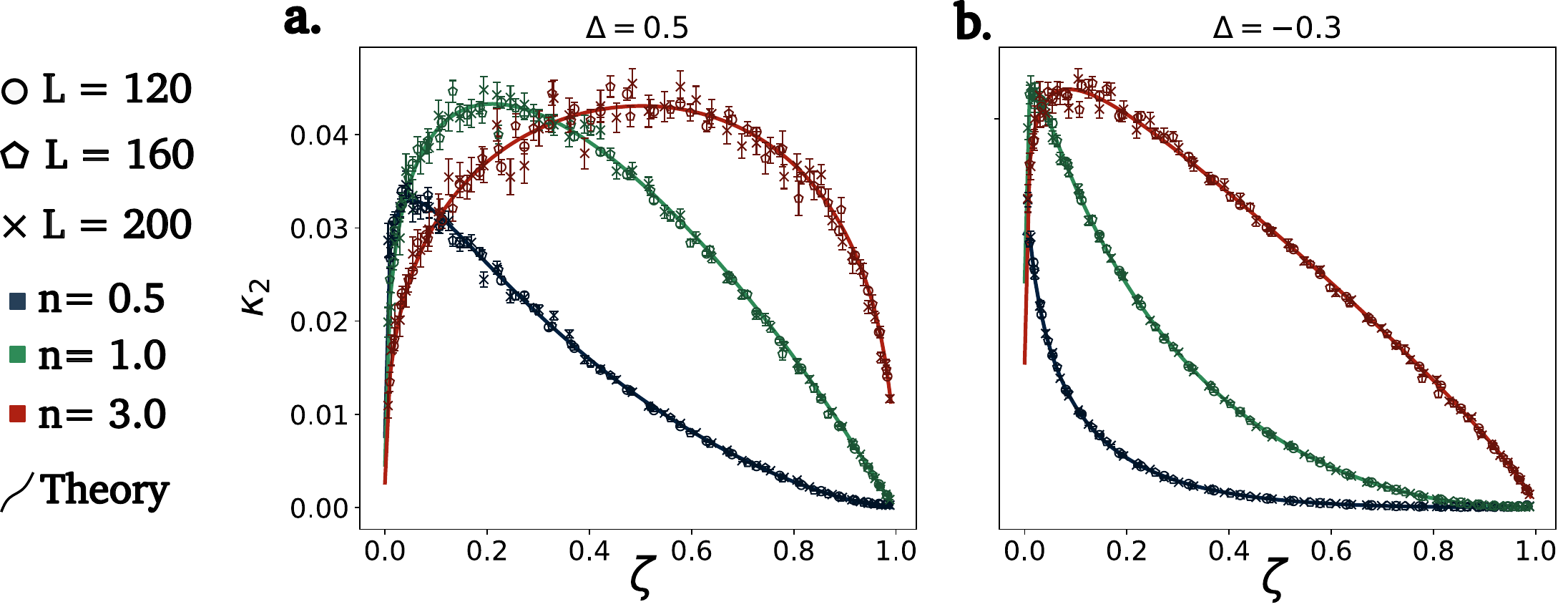}
    \caption{\textbf{Variance $(\kappa_2)$ of MIE versus cross ratio for $\Delta \neq 0$.} Markers show numerical results for the variance of MIE the XXZ chain for the setup in Fig~\ref{Fig1}, and solid curves are theoretical predictions from Eq.~\eqref{cumulantfinalexpression}. 
    (a) $\Delta = 0.5$ (b) $\Delta = -0.3$.}
    \label{plots2}
\end{figure*}
\subsection{Disorder Induced Entanglement}
As an additional proof of concept of equivalency between averaging over lattice measurement outcomes and over conformal boundary conditions at low energies, we briefly analyze entanglement induced by uncorrelated quenched (impurity-like) disorder, not weighted by  Born probabilities. In particular, we define the \textit{disorder induced entanglement} (DIE) as
\begin{equation}
    \mathrm{DIE}^{(n)}(A)=\frac{1}{\mathcal N}\sum_\m S^{(n)}_{\m,A},
\end{equation}
where $\mathcal N$ is the normalization from the number of measurement outcomes in region $B$ of the lattice. The difference from MIE lies in how we weigh measurement outcomes with the disorder average being uniformly distributed over all outcomes. Quenched disorder related phenomena have a rich history in statistical physics. In fact, the tools used to understand measurement related disorder in modern literature are closely related to the ones that were used to understand quenched uncorrelated disorder. For example, both problems require the replica trick to average over the randomness, a trick that introduces $Q$($=nk+1$ in this work) copies (``replicas'') of the systems. One then must take the replica limit at the end of the calculation in order to recover the quantity of interest. The crucial difference between the two disorders lies in how one takes this limit: the inherent Born-weighted randomness of measurements introduced an additional replica which leads to the replica limit $Q \ra 1$ (equivalently, $k\ra0$ like in Eq.~\eqref{avgmiereplica}) \cite{PhysRevB.101.104301, PhysRevB.101.104302}, while quenched disorder is studied via the limit $Q\ra 0$ since the average is effectively over the uniform distribution. These different limits generally produce different universal behavior \cite{jian2023measurementinducedentanglementtransitionsquantum, PhysRevB.108.104203, patil2024highly}, making the disorder case a useful benchmark of our framework. 

Since the only change lies in how one weighs the outcomes, our prescription (from Eq.~\eqref{averagemieprobinterp}) implies that the disorder-induced entanglement, denoted $\mathrm{DIE}$, is the average of forced MIE over the \textit{uniform} distribution, that is
\begin{equation}
\label{DIEprobinterp}
   \mathrm{DIE}^{(n)}= \int_{0}^{2\pi} \frac{d\delta_{\vphi}}{2\pi}\ \mathrm{MIE^{(n)}_F}(\delta_{\vphi}),
\end{equation}
where $\mathrm{MIE_F^n}(\delta_{\vphi})$ is given in Eq.~\eqref{mief}. The same result follows from repeating the replica analysis of Sec~\ref{sectionreplicacalculation}, where one does not need an additional replica coming from the Born-rule. To contrast DIE with MIE, consider the small cross-ratio regime $\zeta\to0$. Using a similar analysis of asymptotics from Eq.~\eqref{asymptoticint} but with $p(\delta_{\vphi})=1/(2\pi)$, we obtain (up to overall constants)
\begin{align}
    \mathrm{DIE}^{(n)}&\underset{\zeta\to 0}{\sim}  
      \int_{0}^{\pi}\! d\delta_{\varphi}\;
     \left[\frac{
        e^{-\frac{2gn\pi^2}{h}\!\left(1-\frac{\delta_{\varphi}}{\pi}\right)}-n
                  e^{-\frac{2g\pi^2}{h}\!\left(1-\frac{\delta_{\varphi}}{\pi}\right)}}{1-n}\right] \notag\\
                  &\underset{\zeta\to 0}{\sim}  \frac{1}{\log(1/\zeta)}\;\;\forall\;\;n,
\end{align}
Thus, the universal behavior of DIE at small $\zeta$ differs starkly from MIE with the former crucially scaling independently of the Rényi index $n$ and Luttinger parameter $g$. Finally, we check Eq.~\eqref{DIEprobinterp} by performing a numerics on the XX chain, where we find good agreement between the two (see Appendix~\ref{dienumerics}). 

\subsection{Universal Distribution of MIE}
\label{universaldistmie}
The $\zeta\ra 0$ scaling form \eqref{scalingcumulants} (where $l=1$) of MIE stand in sharp contrast to what one would expect when considering typical configurations (see Eq.~\eqref{mietypical})—namely the conformal boundary condition $\delta_{\vphi}=0$ (see Eq.~\eqref{pdeltaphi})—for which the forced MIE $(n\mathrel{\underset{\sim}{>}} 1)$ scales as $\sim \zeta^{2g}$ \cite{Rajabpour_2016, khanna2025measurementinducedentanglementconformalfield}. To understand the origin of this discrepancy and to further probe the behavior of MIE, we now study the full distribution of the post-measurement Von-Neumann entropy $S_\m$. We begin by noting that the moment-generating function equals the Laplace transform of $p(S_\m)$, i.e., 
\begin{equation}
   \mathbb{E}[e^{-sS_{\m}}]  = \int_{0}^{\infty} dS_{\m}\;p(S_{\m}) e^{-sS_\m} =\mc L[p(S_\m)](s),
\end{equation}
where $p(S_\m)$ is the distribution of $S_\m$, and we have used that $S_\m\geq 0$. We know from Section~\ref{bornaveragingandcumulants} that the moment generating function of MIE can also be written as an average over conformal boundary conditions, that is,
\begin{equation}
     \mathbb{E}[e^{-sS_{\m}}] =\mc L[p(S_\m)](s)= \int_{0}^{2\pi}d\delta_\varphi\; p(\delta_{\vphi}) \exp\left[-s\mief(\delta_\vphi)\right].
\end{equation}
Taking the inverse Laplace transform then returns
\begin{equation}
    p(S_\m) =  \int_{0}^{2\pi}d\delta_\vphi\; p(\delta_{\vphi}) \mc L^{-1}\{\exp\left[-s\mief(\delta_\vphi)\right]\}=\int_0^{2\pi}d\delta_{\vphi} \;p(\delta_{\vphi})\;\delta(S_\m - \mief(\delta_\vphi)).
\end{equation}
In other words, probability for a given value $S_\m$ to occur is simply the ``sum over probabilities'' of the conformal boundary conditions $\delta_\vphi$ for which the forced MIE equals $S_\m$. Further simplifying the above, one arrives at
\begin{equation}
\label{miedist}
\boxed{
    p(S_\m) = \sum_{\{S_\m = \mief(\delta_{\vphi_{i}})\}} \frac{p(\delta_{\vphi_i})}{|\mief'(\delta_{\vphi_i})|}},
\end{equation}
where $\mief'(\delta_{\vphi_i}) = \left.\tfrac{d\mief}{d\delta_{\vphi}}\right|_{\delta{\vphi}=\delta_{\vphi_i}}$ is assumed to be non-zero. Using $\mief$ from \eqref{mief} we evaluate the above numerically and find that this agrees well with the distribution of MIE obtained from numerical analysis performed on the XX chain (see Fig.~\ref{miedistplots}). The figures illustrate that MIE has a bi-modal distribution with divergences at \textit{both} its ends. This is a signature of the derivative of $\mief(\delta_{\vphi})$ vanishing at $\delta_{\vphi}=\{0,\pi\}$, points that correspond to minimum and maximum values of $\mief(\delta_{\vphi})$ respectively, thereby causing to \eqref{miedist} diverge.
The obtained distribution is rather non trivial and suggests that upon partial  measurement of the critical state, there is a significant chance of generating not only the minimum but also the maximum entanglement that one could possibly obtain in the leftover unmeasured state. To get a better understanding, we briefly discuss how the right tail diverges, in the $\zeta\ra0$ regime which is when the unmeasured regions are maximally separated and the winding contributions to MIE dominate. \\

The divergence in the distribution at the right end comes from the vanishing of $|\mief'(\delta_{\vphi})|$ at $\delta_{\vphi}=\pi$. Here we find that $\mathrm{MIE^{(n)}_F}(\delta_{\vphi}=\pi)\underset{\zeta\ra 0}{\approx} \log 2$. In other words, when the unmeasured part of the system can be approximated as having individual qubits that are maximally separated, one can generate a Bell pair between them by forcing $\delta_{\vphi}=\pi$. A diverging tail at this point is therefore interesting and worth analyzing. For this we begin by introducing a small parameter $\epsilon$ and evaluate the distribution at the neighborhood points $\delta_{\vphi_{0,1}}=\pi\pm\epsilon$. The forced MIE at these points is
\begin{align}
    \mief(\delta_{\vphi}=\pi\pm\epsilon) &= \lim_{n\ra 1}\frac{1}{1-n}\left[\log\left(1+\zeta^{\mp \frac{2ng\epsilon}{\pi}}\right)-n\log\left(1+\zeta^{\mp \frac{2g\epsilon}{\pi}}\right)\right]\\
    & = \log 2-\frac{g^2\log^2\zeta}
    {2\pi^2}\epsilon^2 + \mc O(\epsilon^4),
    \label{epsilonmief}
\end{align}
from which the roots of Eq.~\eqref{miedist} follow as
\begin{equation}
\label{rootsmieflog2}
   \epsilon  \frac{g}{\pi}\log \zeta= \pm \sqrt{2\Delta S_\m},
\end{equation}
where we have defined $\Delta S_\m = \log 2 - S_\m$. Using Eqs.~\eqref{pdeltaphi} and \eqref{miedist}, the distribution in the vicinity of $\delta_{\vphi}=\pi$ can therefore be written as
\begin{equation}
   p(S_\m) \underset{\substack{S_m\ra\log 2^{-}\\ \zeta\ra 0}}{\approx} \sqrt{\frac{g}{2\pi h(\zeta)}}\frac{\pi}{g\log(1/\zeta)}\frac{1 }{\sqrt{2\Delta S_\m}}\left[ \sum_{l\in\mathbb{Z}}\zeta^{2g(l+\frac{1}{2}+\frac{\epsilon}{2\pi})^2} + \zeta^{2g(l+\frac{1}{2}-\frac{\epsilon}{2\pi })^2}\right],
\end{equation}
where the derivative $|\mief'|_{\delta_{\vphi_{0,1}}} = (g/\pi)\log(1/\zeta)\sqrt{2\Delta S_\m}$ follows directly from Eqs.~\eqref{epsilonmief} and \eqref{rootsmieflog2}. Expanding $h(\zeta)=\pi^2/\log(1/\zeta)+\dots$ as $\zeta\ra 0$ and rewriting the result entirely in terms of $S_\m$, we obtain the leading universal behavior
\begin{equation}
\boxed{
\label{scalingrighttail}
    p(S_\m)\underset{\substack{S_m\ra\log 2^{-}\\ \zeta\ra 0}}{\sim} \frac{\zeta^{g/2}}{\sqrt{\log(1/\zeta)}}\frac{1}{\sqrt{\Delta S_\m}}},
\end{equation}
where unimportant factors have been dropped. Thus, upon approaching the distribution's right edge, we see that it exhibits a heavy tail with a square-root divergence $p(S_\m)\propto 1/\sqrt{\Delta S_\m}$ with a pre-factor that vanishes in the limit $\zeta\ra0$. Physically, this indicates a vanishingly small yet finite probability of producing near maximal entanglement by measuring the critical ground state, suggesting that the critical state functions as a ``quantum wire'' in close analogy with SPT phases \cite{PhysRevLett.108.240505}. Finally, notice that the pre-factor in Eq.~\eqref{scalingrighttail} coincides with the scaling of the cumulants $\kappa_l$ in Eq.~\eqref{scalingcumulants} for $n=1$. Indeed, using the above distribution for the right tail, we get
\begin{equation}
   \overline{S_\m^l} \underset{\zeta\ra 0}{\sim}\int_{\log2 -\epsilon}^{\log 2} dS_\m \;  p(S_\m) S^l_\m=(\log2)^l\frac{\zeta^{g/2}}{\sqrt{\log(1/\zeta)}}\int_0^{\epsilon} \frac{dS'}{\sqrt{S'}}\sim \frac{\zeta^{g/2}}{\sqrt{\log(1/\zeta)}}.
\end{equation}
In other words, the right tail near $S_\m=\log2$ controls the leading scaling of the moments, and hence of the cumulants. This shows that MIE and its cumulants are dominated in the $\zeta \ra 0 $ limit by rare large entanglement boundary conditions. However, away from this limit, other boundary conditions start to contribute smoothly. In particular, the $\zeta \ra 1$ regime must cross-over to the usual Cardy-Calabrese \cite{Pasquale_Calabrese_2004} scaling of entanglement, which is certainly not controlled by these rare boundary conditions that contribute with $O(1)$ entanglement.

\section{Conclusion}
We have derived exact expressions for the higher cumulants of MIE, as well as its full distribution, for TLLs. Although the MIE and its cumulants are \textit{a priori} defined on the lattice, they are fully determined in the low-energy limit, where they can be expressed as averages of the forced MIE over boundary conditions labeled by $\delta_\varphi \in [0,2\pi)$, weighted by a probability measure fixed by the partition function of the compact boson on a cylinder with Dirichlet boundary conditions specified by $\delta_\varphi$. As a result, both the cumulants and the full distribution of MIE are universal and conformally invariant, even though some microscopic measurement outcomes are known to break conformal invariance. This suggests that measurement outcomes that break conformal invariance make only a negligible contribution to MIE and its cumulants. As a corollary of microscopic averaging being equivalent to averaging over conformal boundary conditions, we also obtain the DIE, in which measurement outcomes behave as uncorrelated quenched (impurity-like) disorder and are not weighted by Born probabilities but are averaged over uniformly. Finally, we obtain scaling forms for these quantities in the small cross-ratio $(\zeta\ra 0)$ limit where the unmeasured regions are maximally separated. For sufficiently large Rényi index, we find that the entire hierarchy of cumulants of MIE exhibit a universal scaling form, $\zeta^{g/2}/\sqrt{\log(1/\zeta)}$, while for DIE, we find $1/\log(1/\zeta)$ for all Rényi indices. In the same limit, the distribution of MIE is supported on $(0,\log 2)$ and displays heavy tails at both ends, with the weight near $S_{\m}=\log 2$ vanishing as $\zeta\to0$. We find that the tail at $S_\m = \log 2$ diverges as $1/\sqrt{\log 2 - S_\m}$, with a vanishing weight of $\zeta^{g/2}/\sqrt{\log(1/\zeta)}$, indicating a vanishingly small yet finite probability of inducing a Bell-pair across the system through measurements. We also show that in this limit, it is this tail that controls the leading scaling of all the cumulants of MIE for high enough Rényi index.

A natural follow-up of our work would be to further understand this problem for other conformal field theories such as Ising, Potts, and tri-critical Ising. In particular, it is to be checked whether the interpretation of averaging over microscopic Born-probabilities always leads to a Born-average over conformal boundary conditions of those theories. If so, such a result would pave the way for a complete understanding of how measurements affect critical quantum states. Finally, it would also be interesting to study the effects of noise on our setup. 

{\bf Acknowledgements.} We thank Andreas W. W. Ludwig and Sara Murciano for insightful discussions on related topics. This work was supported by the Swiss National Science Foundation (grant 10008234).\\





\begin{appendix}
\section{$\zeta\ra 0 $ Asymptotic Analysis of Cumulants of MIE}
\label{cumulantsappendix}
In this section we analyze the scaling of the cumulants of MIE in the $\zeta\to 0$ limit. It is convenient to first extract the leading behavior of the moments, which—as we show—controls the cumulants in this regime. Using the Born-averaging prescription from the main text, the $l$-th moment is
\begin{align}
  \overline{(S^{(n)}_{\mathbf{m},A})^l}
  = \int_{0}^{2\pi}\! d\delta_{\varphi}\; p(\delta_{\varphi}) \bigl[\mathrm{MIE}^{(n)}_{\mathrm F}(\delta_{\varphi})\bigr]^l,
\end{align}
where $\mathrm{MIE}^{(n)}_{\mathrm F}(\delta_{\varphi})$ and $p(\delta_{\varphi})$ are given in Eqs.~\eqref{mief} and \eqref{pdeltaphi}. Expanding both functions yields
\begin{equation}
  \overline{(S^{(n)}_{\mathbf{m},A})^l}
  = \sqrt{\frac{g}{2\pi h}}\;\frac{1}{(1-n)^l}\!
    \int_{0}^{2\pi}\! d\delta_{\varphi}\;
    \sum_{v\in\mathbb{Z}} q_1^{\,g\!\left(v+\frac{\delta_{\varphi}}{2\pi}\right)^2}
    \Biggl[
      \log\!\frac{\eta(q_1)^n}{\eta(q_n)}
      + \log\!\frac{\sum_{w\in\mathbb{Z}} q_n^{\,g\!\left(w+\frac{\delta_{\varphi}}{2\pi}\right)^2}}
                       {\Bigl(\sum_{l\in\mathbb{Z}} q_1^{\,g\!\left(l+\frac{\delta_{\varphi}}{2\pi}\right)^2}\Bigr)^{\!n}}
    \Biggr]^{\!l},
\end{equation}
with $q_n = e^{-2\pi^2/(h/n)}$. Using Eq.~\eqref{hzeta}, $\zeta\to 0$ implies $h\ra 0$ and $q_n\to 0$ with $h\sim \pi^2/\log(1/\zeta)$ and $q_n \sim \zeta^{2n}$. Consequently, the Dedekind-eta ratio part of the integral is indepdendent of $\delta_\vphi$ and gives a contribution $\log\!\bigl[\eta(q_1)^n/\eta(q_n)\bigr]\sim \zeta^{2n}$ that is sub-leading compared to the winding-sum contribution that (as we shall see) will produce a $\zeta^{g/2}$-like scaling. Due to this, we only keep the leading contributions from the winding sum, leading to
\begin{equation}
  \overline{(S^{(n)}_{\mathbf{m},A})^l}
  \underset{\zeta\to 0}{\sim}
  \sqrt{\frac{1}{h}}\;\frac{1}{(1-n)^l}
  \int_{-\infty}^{\infty}\! d\delta_{\varphi}\;
  e^{-\frac{g\,\delta_{\varphi}^2}{2h}}
  \left[
    \log\!\left(
      \frac{1+e^{-\frac{2\pi^2ng}{h}\left(1+\frac{\delta_{\varphi}}{\pi}\right)}
              +e^{-\frac{2\pi^2ng}{h}\left(1-\frac{\delta_{\varphi}}{\pi}\right)}+\dots}
           {\bigl(1+e^{-\frac{2\pi^2g}{h}\left(1+\frac{\delta_{\varphi}}{\pi}\right)}
                    +e^{-\frac{2\pi^2g}{h}\left(1-\frac{\delta_{\varphi}}{\pi}\right)}+\dots\bigr)^{\!n}}
    \right)
  \right]^{\!l},
\end{equation}
where we have extended $p(\delta_{\varphi})$ to the real line. It is then not too hard to check that for small $\zeta$, one can restrict the integrand to the domain $(-\pi,\pi)$ since the remaining region has an exponentially suppressed contribution at $\mathcal O(e^{-2\pi^2 g/h})$ that we ignore. For $|\delta_{\varphi}|<\pi$, we have $e^{-\frac{2\pi^2g}{h}(1+\delta_{\varphi}/\pi)} \allowbreak \ll e^{-\frac{2\pi^2 g}{h}(1-\delta_{\varphi}/\pi)}\ll  1$ as $\zeta\to 0$ and so we have
\begin{equation}
\label{asymptoticintcontrolling}
  \overline{(S^{(n)}_{\mathbf{m},A})^l}
  \underset{\zeta\to 0}{\sim}
  \sqrt{\frac{1}{h}}
  \int_{0}^{\pi}\! d\delta_{\varphi}\;
  e^{-\frac{g\,\delta_{\varphi}^2}{2h}}
  \left[
    \frac{e^{-\frac{2gn\pi^2}{h}(1-\frac{\delta_{\varphi}}{\pi})}
          - n\,e^{-\frac{2g\pi^2 }{h}(1-\frac{\delta_{\varphi}}{\pi})}}{1-n}
  \right]^{\!l}, 
\end{equation}
using $\log(1+\epsilon)\simeq \epsilon$ for $\epsilon\ll 1$ and even-ness of the integral. The above is the final integral that controls the behavior of cumulants, as was discussed for MIE (see Eq.\eqref{asymptoticint}) in the main text. Let us now analyze various cases of the Rényi index $n$:
\begin{itemize}
    \item For $n<1$, we have 
    \begin{equation}
          \overline{(S^{(n)}_{\mathbf{m},A})^l}
  \underset{\zeta\to 0}{\sim}
  \sqrt{\frac{1}{h}}
  \int_{0}^{\pi}\! d\delta_{\varphi}\;
  e^{-\frac{g\,\delta_{\varphi}^2}{2h}} e^{-\frac{2gnl\pi^2}{h}(1-\frac{\delta_{\varphi}}{\pi})}.
    \end{equation}
where we have dropped unimportant $n$-dependent factors. Completing the square in the integrand gives 
\begin{equation}
   \overline{(S^{(n)}_{\mathbf{m},A})^l}\underset{\zeta\to 0}{\sim} \frac{e^{-\frac{g}{2h}4\pi^2nl(1-nl)}}{\sqrt{h}}\int_0^\pi d\delta_\varphi e^{-\frac{g}{2h}(\delta_\varphi - 2\pi nl)^2}.
\end{equation}
For small $h$, or equivalently, small $\zeta$, one can use the saddle point method to estimate how the above integral scales.\\
For $nl>1/2$, the saddle is outside the domain $[0,\pi]$ and so the integrand attains its minima at $\delta_\varphi = \pi$. We can then estimate $\int_0^{\pi}e^{-(g/2h)(\delta_\varphi - 2\pi nl )^2}\sim \int_{-\infty}^{\pi} e^{-(g/2h)(\delta_\varphi - 2\pi n l )^2}$ at small $h$ to be scaling as $\sim h e^{-g\pi^2(1-2nl)^2/2h}$. For $nl<1/2$ on the other hand, the saddle lies in the interior of the domain $[0,\pi]$, and the integral then scales as $\int_{-\infty}^{\infty} e^{-(g/2h)(\delta_\varphi - 2\pi n l )^2}\newline\sim \sqrt{h}$. At $nl=1/2$, the end-point of the domain, $\pi$, is exactly at the maxima of the gaussian and so the integral scales as $\int_{-\infty}^0 dt\; e^{-(g/2h)t^2} \sim \sqrt{h}$. Putting this all together, we have
\begin{equation}
\label{scalingcumulantsapp}
\overline{(S^{(n)}_{\mathbf{m},A})^l}
\underset{\zeta \to 0}{\sim}
\begin{cases}
\displaystyle \dfrac{\zeta^{g/2}}{\sqrt{\log(1/\zeta)}} & n > \tfrac{1}{2l}, \\[1.1ex]
\displaystyle \zeta^{g/2} & n = \tfrac{1}{2l}, \\[0.8ex]
\displaystyle \zeta^{2gnl(1-nl)} & 0 < n < \tfrac{1}{2l},
\end{cases}
\end{equation}
where, as discussed in the main text (see Sec.~\ref{universalscalingcumulantszeta0}), the transitions in qualitative behavior of the scalings correspond to how the limits of the  integral Eq.~\eqref{asymptoticintcontrolling} are positioned w.r.t the mean when re-cast as a pure Gaussian. 
\item For $n=1$, we can take the replica limit $n\ra 1$ in Eq.~\eqref{asymptoticintcontrolling}, giving

\begin{equation}
\overline{(S_{\mathbf{m},A})^l}
\underset{\zeta \to 0}{\sim} \frac{1}{\sqrt{h}}\int_0^{\pi} d\delta_{\vphi}\;e^{-\frac{g\delta_{\vphi}^2}{2h}}e^{-\frac{2\pi^2 g l}{h}(1-\frac{\delta_\vphi}{\pi})}\left[\frac{2\pi^2 g}{h}\left(1-\frac{\delta_\vphi}{\pi}\right)+1\right]^l.
\end{equation}
As before, we can complete the square of the integrand, leading to
\begin{equation}
       \overline{(S_{\mathbf{m},A})^l}\underset{\zeta\to 0}{\sim} \frac{e^{-\frac{g}{2h}4\pi^2l(1-l)}}{\sqrt{h}}\int_0^\pi d\delta_\varphi e^{-\frac{g}{2h}(\delta_\varphi - 2\pi l)^2}\left[\frac{2\pi^2 g}{h}\left(1-\frac{\delta_\vphi}{\pi}\right)+1\right]^l.
\end{equation}
For $l>1/2$, the saddle is outside the domain $[0,\pi]$ as was the case for $nl<1/2$ previously. However, here one has to be careful with the algebraic pre-factors in the integrand that can play a role. The cleanest way to extract scaling is to change variables as $\delta_\varphi = \pi - h\delta_\varphi'$ and re-write the integral above. Doing so gives  $e^{-g\pi^2(1-2l)^2/2h} \;h \times \newline\int_{-\infty}^0 d\delta_\varphi' e^{-(gh/2)\delta_\varphi'^2} e^{-g\delta_\varphi'(2l-1)}(2\pi g\delta_\varphi'/\pi + 1)^l$, where for small $h$, the leftover integral is $O(1)$ in $h$. The pre-factor therefore does not contribute to scaling in this case. Next, for $l<1/2$, the peak of the gaussian is within the domain $[0,\pi]$. A change of variables to $\delta_\varphi = 2\pi l +\sqrt{h}s$ shows that while the gaussian gives a scaling of $\sim \sqrt{h}$, the pre-factor contributes a scaling of $\sim h^{-l}$. This leads to the integral scaling as $\sim h^{1/2 - l}$. Finally, for $l=1/2$, the same change of variables leads to a scaling of $\sim \sqrt{h}h^{-l/2}$, where the pre-factor contributes with $\sim h^{-l/2}$. Putting these together, one finds
\begin{equation}
\label{scalingcumulantsn1}
\overline{(S_{\mathbf{m},A})^l}
\underset{\zeta \to 0}{\sim}
\begin{cases}
\displaystyle \dfrac{\zeta^{g/2}}{\sqrt{\log(1/\zeta)}} & l > \tfrac{1}{2}, \\[1.1ex]
\displaystyle \log(1/\zeta)^{1/4}\;\zeta^{g/2} & l = \tfrac{1}{2}, \\[0.8ex]
\displaystyle \log(1/\zeta)^l \; \zeta^{2gl(1-l)} & 0 < l < \tfrac{1}{2},
\end{cases}
\end{equation}
\item Finally, for $n>1$ the leading scaling is given by the linear term in $n$ in Eq.~\eqref{asymptoticintcontrolling}. The analysis in this case is identical as done for the case of $n<1$, leading to 
\begin{equation}
\label{scalingcumulantsappn1}
\overline{(S^{(n)}_{\mathbf{m},A})^l}
\underset{\zeta \to 0}{\sim}
\begin{cases}
\displaystyle \dfrac{\zeta^{g/2}}{\sqrt{\log(1/\zeta)}} & l > \tfrac{1}{2}, \\[1.1ex]
\displaystyle \zeta^{g/2} & l = \tfrac{1}{2}, \\[0.8ex]
\displaystyle \zeta^{2gl(1-l)} & 0 < l < \tfrac{1}{2},
\end{cases}
\end{equation}
\end{itemize}
Putting it all together, we get Eq.~\eqref{scalingcumulants} in the main text, where we focus only on the integer cumulants. 
\section{Analytic Continuation of $\mc W^{(n)}_{k_1,k_2}$}
\label{analyticcontwn}
In this section we present a brief analytic continuation of $\mc W^{(n)}_{k_1,k_2}$ from Eq.~\eqref{generalizedwindingqn} to Eq.~\eqref{generalizedwinding}. The analytic continuation we present here closely follows that of a similar winding function obtained in Ref.~\cite{Zhou_2016} in a different context. We start by re-writing \eqref{generalizedwindingqn} the above using the Poisson re-summation formula \cite{7a6ba7cf-6d4a-3566-be69-751572b726ce}, giving
\begin{equation}
    \mc W^{(n)}_{k_1,k_2} = \left(\frac{2\pi n g}{h}\right)^{-(k_1+k_2)/2}\sum_{\vec{w}\in\mathbb{Z}^{k_1+k_2}}\tilde q_n^{\vec{w}^T T^{-1}_{k_1+k_2} \vec{w}/(4g)},
\end{equation}
where $\tilde q_n = e^{-\frac{4\pi}{\beta}\frac{h(\zeta)}{n}}$ and 
\begin{equation}
T_{k_1+k_2}^{-1} =
\left[
\begin{array}{ccc|ccc}
n+1 & n  &\cdots & n & n & \dots \\
n   & n+1 &  \cdots & n & n & \cdots \\
\vdots   & \vdots &\ddots& \vdots & \vdots & \ddots \\
\hline
n & n &\dots& 2n & n & \cdots \\
n & n & \dots&n & 2n & \cdots \\
\vdots   & \vdots &\ddots  & \vdots & \vdots & \ddots
\end{array}
\right]_{(k_1+k_2)\times (k_1+k_2)},
\end{equation}
is completely independent of $k_1$ and $k_2$ and contains a top left block of size $k_1\times k_1$ marked above. Analytic continuation is then carried forward by completing the square of the expression $\vec{w}^T T^{-1}_{k_1+k_2} \vec{w}$ and using the Dirac delta function to write
\begin{equation}
   q_n^{\vec{w}^T T^{-1}_{k_1+k_2} \vec{w}/(4g)}=\sum_{\vec w\in\mathbb{Z}^{k_1+k_2}}\int dx\;\exp\left[-\frac{h}{2gn}\sum_{i=1}^{k_1} w_i^2-\frac{h}{2g}\sum_{i=k_1+1}^{k_1+k_2}w_i^2 - \frac{h}{2g}x^2\right]\delta(x-\sum_{i=1}^{k_1+k_2}w_i).
\end{equation}
Expressing Dirac-delta in its Fourier form $\delta(x-a) = \int_{-\infty}^{\infty}\frac{d\vphi}{2\pi} e^{ip(x-a)}$ and integrating over $x$ results in 
\begin{equation}
\begin{split}
    q_n^{\vec{w}^T T^{-1}_{k_1+k_2} \vec{w}/(4g)} &=\sqrt{\frac{2\pi g}{h}}\sum_{\vec w\in\mathbb{Z}^{k_1+k_2}}\int \frac{d\vphi}{2\pi }\;e^{-g\vphi^2/(2h)}\exp\left[-\frac{h}{2gn}\sum_{i=1}^{k_1} w_i^2-\frac{h}{2g}\sum_{i=k_1+1}^{k_1+k_2}w_i^2 - i\vphi\sum_{i=1}^{k_1+k_2}w_i\right]\\
     &= \sqrt{\frac{2\pi g}{h}}\int\frac{d\vphi}{2\pi}e^{-g\vphi^2/(2h)}\left[\sum_{w\in\mathbb Z}\exp\left(-\frac{h}{2gn}w^2 - i\vphi w\right)\right]^{k_1}\left[\sum_{w\in\mathbb Z}\exp\left(-\frac{h}{2g}w^2 - i\vphi w\right)\right]^{k_2}.
    \end{split}
\end{equation}
where in the second line we factorized the sum over $\vec{w}\in\mathbb{Z}^{k_1+k_2}$ to $k_1+k_2$ independent sums over $\mathbb{Z}$. Re-writing these summations using the Poisson re-summation formula \cite{DiFrancesco1997}
\begin{equation}
    \sum_{w\in\mathbb{Z}} \exp\bigg[-\pi a w^2 + 2\pi i b w\bigg] = \frac{1}{\sqrt{a}}\sum_{w\in \mathbb{Z}}\exp\bigg[ - \frac{\pi}{a}(w+b)^2\bigg],
\end{equation}
results exactly in the expression \eqref{generalizedwinding}. 
\section{Numerics}
\label{numericsappendix}
In this section, we detail the numerical methods used to obtain results in Figs.~\ref{plots} and~\ref{miedistplots}. We use exact matrix product state (MPS) calculations on the XX chain which can be mapped to a model of free-fermions at half filling with the Hamiltonian
\begin{equation}
    H = -\sum_i \left(c_i^{\dag}c_{i+1} +\; \text{h.c.}\right)+\mathrm{const.}
\end{equation}
with periodic (antiperiodic) boundary conditions when the total number of fermions is odd (even). The numerics for this case can be done exactly since the ground state is Gaussian and hence is entirely determined by its correlation matrix \cite{ giamarchi_quantum_2003}
\begin{equation}
    C_{ij}=  \langle c_i^{\dag}c_j\rangle = \frac{\sin(\pi n_f(i-j))}{L\sin\frac{\pi(i-j)}{L}},
\end{equation}
where $n_f$ is the fermion-filling factor which is $1/2$ in our case. Charge ($\sigma_z$) measurements are implemented by updating the correlation matrix as per the rules
\begin{equation}
C'_{ij}=\frac{\langle c_a^\dagger c_a\, c_i^\dagger c_j\, c_a^\dagger c_a\rangle}{C_{aa}}
=\begin{cases}
1, & i=j=a,\\[4pt]
C_{ij}-\dfrac{C_{ia}C_{aj}}{C_{aa}}, & i\neq a,\; j\neq a,\\[8pt]
0, & \text{otherwise},
\end{cases}
\tag{A3}
\end{equation}
when we apply the projector $P_1 = c^{\dag}_{a}c_a$ with probability $p_a = C_{aa}$, where $a$ is the measured orbital (site). Similarly, when we apply the projector $P_0 = 1-c_a^{\dag}c_a$ with probability $p_0=1-C_{aa}$, the updated correlation matrix is

\begin{equation}
C'_{ij}=\frac{\langle c_a c_a^\dagger\, c_i^\dagger c_j\, c_a c_a^\dagger\rangle}{1-C_{aa}}
=\begin{cases}
0, & i=j=a,\\[4pt]
C_{ij}+\dfrac{C_{ia}C_{aj}}{1-C_{aa}}, & i\neq a,\; j\neq a,\\[8pt]
0, & \text{otherwise}.
\end{cases}
\tag{A4}
\end{equation}
where multi-particle correlators can be evaluated using Wick’s theorem. The above rules are easy to derive (see Ref.~\cite{PhysRevB.109.195128}). With the above update rules, one can obtain the resultant correlation matrix upon measuring all the sites in region $B$, appropriately sampling them via their respective Born-probabilities. The entanglement entropy of region $A$ is then easily calculated via the obtained correlation matrix \cite{Peschel_2009}. In Fig.~\ref{plots}a and b, the sample sizes range from $2\times 10^3$ to $7.5\times 10^4$, depending on the cross-ratio $\zeta$, with smaller cross-ratios requiring a larger number samples due to a larger measurement record.\\
For the interacting XXZ data shown in Figs.~\ref{plots2}a,b, we compute the ground state using DMRG as implemented in the \texttt{iTensor} library~\cite{itensor,white1992density,white1993density}. We consider chains of length $L=120,160,200$ at $\Delta=-0.3$ and $\Delta=0.5$. After obtaining the MPS ground state, measurement outcomes in the $\sigma_z$ basis are sampled (as per the Born-rule) by applying the corresponding local projectors to the state. The DMRG bond dimension is increased over the course of the sweeps while keeping the truncation error below $10^{-7}$. The largest bond dimensions reached are $(364,312)$ for $L=120$, $(445,379)$ for $L=160$, and $(512,436)$ for $L=200$, where the two entries correspond to $\Delta=(0.5,-0.3)$, respectively. Across the data shown in the main text, the number of samples per point ranges from roughly $2\times 10^2$ to $9\times 10^3$, depending on the cross-ratio $\crat$; smaller values of $\crat$ involve larger measured regions and therefore require more sampling.

\section{Numerics for Disorder Induced Entanglement}

\label{dienumerics}
We benchmark our theoretical prediction for DIE, Eq.~\eqref{DIEprobinterp}, against free-fermion numerics described in the previous section of the Appendix. As shown in Fig.~\ref{plotdie}, the analytic result is in good agreement with the numerical data, with the deviations at small cross-ratios consistent with finite-size effects. The number of samples in the plot below range from $7\times 10^2$ to $1.9 \times 10^4$ depending on the cross-ratio. 
\begin{figure*}[h]
    \centering
    \includegraphics[width=0.7\linewidth]{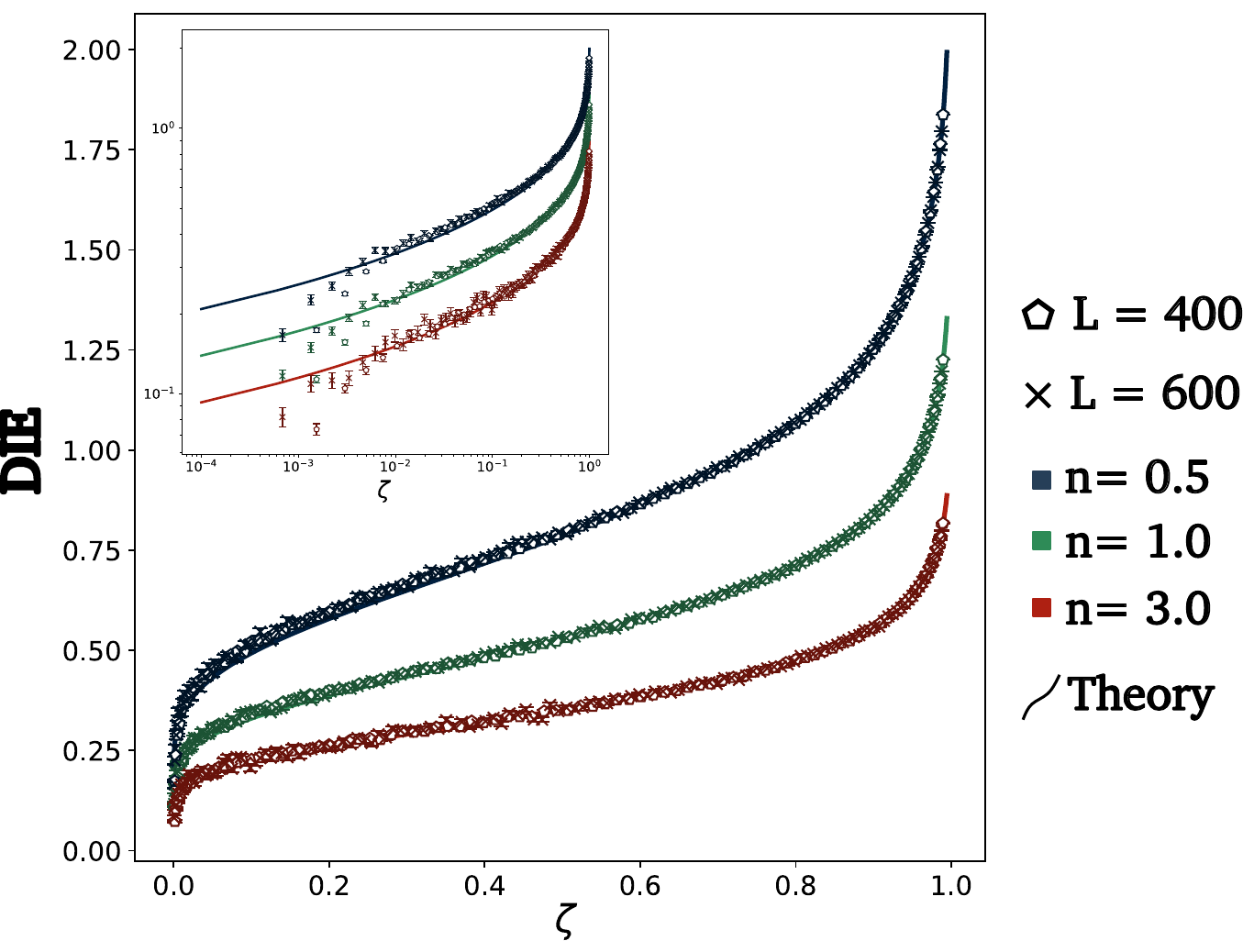}
    \caption{\textbf{Disorder Induced Entanglement (DIE).} Markers show numerical results and solid curves are theoretical predictions.}
    \label{plotdie}
\end{figure*}
\\

\end{appendix}

\bibliography{SciPost_Example_BiBTeX_File.bib}

\end{document}